\documentclass[preprint]{revtex4}
\usepackage[T1]{fontenc}
\usepackage{graphicx}
\usepackage{rotating}
\usepackage{bm}        
\usepackage{amssymb}   
\usepackage{bm}
\usepackage{float}
\usepackage{tikz}

\def\jpca#1#2#3{{\it J.~Phys.~Chem.~{\rm A}}~{\bf #1},\ #2\ (#3)}
\def\jpcc#1#2#3{{\it J.~Phys.~Chem.~{\rm C}}~{\bf #1},\ #2\ (#3)}
\def\jcp#1#2#3{{\it J.~Chem.~Phys.}~{\bf #1},\ #2\ (#3)}

\def\pra#1#2#3{{\it Phys.~Rev.}~A~{\bf #1},\ #2\ (#3)}
\def\prl#1#2#3{{\it Phys.~Rev.~Lett.}~{\bf #1},\ #2\ (#3)}

\def\mp#1#2#3{{\it Mol. Phys.}~{\bf #1},\ #2\ (#3)}
\def\jpb#1#2#3{{\it J. Phys. B: At. Mol. Opt. Phys.} {\bf #1},\ #2\ (#3)}

\def\epjd#1#2#3{{\it Eur.~Phys.~J~{\rm D}}~{\bf #1},\ #2\ (#3)}
\def\pccp#1#2#3{{\it Phys. Chem. Chem. Phys.}~{\bf #1},\ #2\ (#3)}
\def\irpc#1#2#3{{\it Int. Rev. Phys. Chem.}~{\bf #1},\ #2\ (#3)}
\def\jctc#1#2#3{{\it J. Chem. Theor. Comp.}~{\bf #1},\ #2\ (#3)}
\def\ijqc#1#2#3{{\it Int. J. Quant. Chem.}~{\bf #1},\ #2\ (#3)}
\def\njp#1#2#3{{\it New J. Phys.}~{\bf #1},\ #2\ (#3)}
\def\apj#1#2#3{{\it Astrophys. J.}~{\bf #1},\ #2\ (#3)}

\def\k1{k_1}
\def\k2{k_2}
\def\q1{q_1}
\def\q2{q_2}

\def\({\left (}
\def\){\right )}
\def\[{\left [}
\def\]{\right ]}

\newcommand{\beq}{\begin{equation}}
\newcommand{\eeq}{\end{equation}}

\begin{document}
\date{\today}
\flushbottom \draft
\title{Bayesian machine learning for quantum molecular dynamics
}
\author{R. V. Krems\footnote{E-mail address: rkrems@chem.ubc.ca}}
\affiliation{
Department of Chemistry, University of British Columbia, Vancouver, B.C. V6T 1Z1, Canada
}
\begin{abstract}

This article discusses applications of Bayesian machine learning for quantum molecular dynamics. 
One particular formulation of quantum dynamics advocated here is in the form of a machine learning simulator of the Schr\"{o}dinger equation. If combined with the Bayesian statistics, such a simulator allows one to obtain not only the quantum predictions but also the error bars of the dynamical results associated with uncertainties of inputs (such as the potential energy surface or non-adiabatic couplings) into the nuclear Schr\"{o}dinger equation. 
 Instead of viewing atoms as undergoing dynamics on a given potential energy surface, Bayesian machine learning allows one to formulate the problem as the Schr\"{o}dinger equation with a non-parametric distribution of potential energy surfaces that becomes conditioned by the desired dynamical properties (such as the experimental measurements). Machine learning models of the Schr\"{o}dinger equation solutions can identify the sensitivity of the dynamical properties to different parts of the potential surface, the collision energy, angular momentum, external field parameters and basis sets used for the calculations. This can be used to inform the design of efficient quantum dynamics calculations. Machine learning models can also be used to correlate rigorous results with approximate calculations, providing accurate interpolation of exact results.  Finally, there is evidence that it is possible to build Bayesian machine learning models capable of physically extrapolating the solutions of the Schr\"{o}dinger equation. This is particularly valuable as such models could complement common discovery tools to explore physical properties at Hamiltonian parameters not accessible by rigorous quantum calculations or experiments, and potentially be used to accelerate the numerical integration of the nuclear Schr\"{o}dinger equation.

\end{abstract}

\maketitle

\clearpage
\newpage

\section{Introduction}

Machine learning (ML) has become an important tool for physics and chemistry research \cite{general-ML-review}. 
While many of the ML methods date back several decades, the last five to ten years have seen an explosion of high-impact publications applying ML to problems in science. 
This can be attributed to two factors. First, the computer processing power has reached the level making optimization of complex, multi-dimensional non-convex functions feasible, which is necessary for training ML models.  Second, multiple open-source software packages have been developed for applications of ML methods. It has thus become possible to combine traditional quantum dynamics calculations, molecular dynamics simulations  and chemistry experimentation with ML. 
 While ML is already used extensively to assist molecular dynamics simulations for a variety of applications \cite{ML-for-MD-1, ML-for-MD-2, ML-for-MD-3}, density functional theory \cite{ML-for-DFT-1, ML-for-DFT-2, ML-for-DFT-3, ML-for-DFT-4, ML-for-DFT-5, ML-for-DFT-6, ML-for-DFT-7}, the design of chemistry experiments \cite{ML-for-Chemistry-1,ML-for-Chemistry-2,ML-for-Chemistry-3,ML-for-Chemistry-4}, new materials discovery \cite{ML-for-Mat-1, ML-for-Mat-2, ML-for-Mat-3, ML-for-Mat-4} and the prediction of molecular properties \cite{ML-for-MolProp-1, ML-for-MolProp-2}, it is only beginning to have an impact on the research field of quantum reaction dynamics of molecules. 
The purpose of this article is to describe what can be gained from combining ML with quantum dynamics calculations aimed at understanding the microscopic reactions of molecules. 
 
 In general, there are three types of ML algorithms:  supervised learning, unsupervised learning, and reinforcement learning.  The goal of supervised learning is to build a model of input $\leftrightarrow$ output relationships, given a finite number of examples of such relationships. Fitting a potential energy surface for a polyatomic molecule with an artificial neural network (NN) is an example of supervised learning \cite{NNs-for-PES-4}. In this example, the input variables could be the atomic coordinates $\bm x = [x_1, x_2, ..., x_N]^\top$, the output $y$ is the potential energy of the molecule and the NN provides a model $y(\bm x)$ of the global surface given a finite number of {\it ab initio} energy points $y (\bm x)$ in the configuration space. Unsupervised learning aims to analyze the properties of given data without any prior information.  Principal component analysis is one example of unsupervised learning.  Reinforcement learning is closely aligned with optimal control theory. Accurate input $\leftrightarrow$ output relationships are not needed for reinforcement learning. Rather, the purpose of reinforcement learning is to build a model that maps various input states onto desired outputs through an iterative process guided by some reward policy. An example of reinforcement learning is the inverse scattering problem aiming to construct a ML model of a potential energy surface that yields quantum dynamics results in full agreement with experimental results \cite{rodrigo-bo}. 
In the present article, most of the focus is on supervised and reinforcement learning, as well as optimization with machine learning. 

Much of ML is done with artificial neural networks. NNs provide flexible models of the input $\leftrightarrow$ output relationships, often in the form of an analytic, deterministic non-linear fit of $y(\bm x)$. There are also stochastic NNs, such as Boltzmann machines \cite{general-ML-review}, which can be used to model probability distributions, and Bayesian NNs \cite{BML}, which provide probabilistic predictions.  In general, NNs aim to {\it fit} the input $\leftrightarrow$ output relationships. 
In general, in order to be flexible and accurate, NNs must be complex. This complexity requires a large number of input $\leftrightarrow$ output relationships to train NNs. 
Another class of ML methods relies on kernels \cite{kernel-methods-in-ML}. A few examples of such methods include nonlinear support vector machines, kernel ridge regression or Gaussian process (GP) regression. Kernel regression methods often {\it use} the input $\leftrightarrow$ output relationships directly. For example, given $n$ values of $\bm y = [y_1, y_2, .., y_n]^\top$ at various values of $\bm x$, GP regression provides a prediction model in the form of a linear combination of the given outputs $y_i$. 
For this reason, the numerical difficulty of constructing accurate kernel-based ML models scales between ${\cal O}(n^2)$ and  ${\cal O}(n^3)$, where $n$ is the number of training points (if all training points are used simultaneously), which becomes more expensive than training NNs as $n \rightarrow \infty$.  On the other hand, kernel methods
generally require fewer input $\leftrightarrow$ output relationships to make accurate predictions than NNs. In quantum dynamics, producing information is time-consuming because it requires solving the Schr\"{o}dinger equation.  The limiting step in applications of ML to quantum dynamics will thus often be to obtain the input $\leftrightarrow$ output relationships. Therefore, 
one of the goals of the present article is to argue that
kernel methods are more suitable for a productive symbiosis between ML and quantum dynamics than ML methods based on NNs, as they can obtain more information from fewer  input $\leftrightarrow$ output relationships. This point will be illustrated by multiple examples throughout this article. 

This article also argues that the research field of molecular dynamics can benefit from Bayesian optimization (BO).  
This optimization technique relies on probabilistic ML models, which provide not only a prediction but also an uncertainty of a prediction. Both the prediction itself and the prediction uncertainty are exploited to build a powerful algorithm, which can be used to raise and answer questions regarding quantum molecular dynamics generally  considered unfeasible. While Bayesian ML is generally difficult, GP regression offers a straightforward way to implement BO. I present examples illustrating the power and efficiency of BO in application to quantum reactive scattering problems and discuss potential applications of BO in quantum dynamics research.   Bayesian inference has already been used with much success in classical molecular dynamics, as exemplified by Refs. \cite{bayesian-calibration-1,bayesian-calibration-2,bayesian-calibration-3,bayesian-calibration-4,bayesian-calibration-5,bayesian-calibration-6,bayesian-calibration-7}. 

\subsection{Goals of this article}

The main goals of this article are to illustrate by examples that
\begin{itemize}
\item A combination of ML with quantum dynamics calculations can be used to provide improved quantum dynamics results; 
\item Combining ML with quantum dynamics calculations can be used to address new questions generally considered unfeasible; 
\item ML provides a way to automate many calculations (such as fitting PES) often done manually;
\end{itemize}
and to propose new applications of ML for problems in molecular dynamics, such as, 
\begin{itemize}
\item The possibility of solving the inverse scattering problem; 
\item The possibility of reducing the dimensionality of the relevant Hilbert space for time-independent quantum dynamics calculations; 
\item Efficient interpolation of accurate quantum results;
\item Efficient basis set correction; 
\item Extrapolation of quantum results beyond the range of Hamiltonian parameters accessible by the numerical integration of the Schr\"{o}dinger equation;
\end{itemize}

\subsection{Article organization}

This article is not intended as a review of ML methods. I discuss the ML methods only for the purpose of the goals outlined in Section I.A above. 
Most of the results presented are based on GP regression and BO using Gaussian processes. Therefore, I begin Section II by describing the Bayesian framework for ML. 
I then introduce Gaussian Processes as a limit of a Bayesian Neural Network and describe the algorithm of Bayesian optimization.  Section III summarizes some of the most important problems in quantum reaction dynamics that could benefit from ML. 
Section IV presents examples illustrating the points outlined in Section I.A, while Section V discusses possible future applications of ML in quantum molecular dynamics.

\section{Supervised learning within Bayesian Framework}

This section discusses three key ingredients necessary to make the points outlined in Section I.A and to understand the examples presented in the following sections. 
First, I describe the general concept of Bayesian ML. I then describe Gaussian process regression as a supervised learning method. To make the connection with neural networks and introduce Bayesian ML, I present GPs as a limit of a Bayesian NN. 
Second, I describe the algorithm of Bayesian optimization based on GP regression. Third, I describe
how GP regression can be used to build models capable of predicting physical properties of complex quantum systems outside the range of the training data. 
The discussion in the following sections refers back to the equations presented here. 

\subsection{Gaussian Process as a limit of a Bayesian Neural Network}

Consider an unknown, multi-dimensional function $y(\bm x)$, where $\bm x = [x_1, x_2, ..., x_N]^\top$, and $N$ is the number of independent variables. 
Each of the independent variables $x_i$ spans a certain range $x_i \in \left [ x_i^{\rm min}, x_i^{\rm max} \right ]$. 
A typical supervised learning problem begins with $n$ known values of this function $\bm y = [y_1, y_2, ..., y_n]^\top$ at $n$ different points of this $N$-dimensional space.  
These values are referred to as `training points'. 
The goal is to build a model of $y(\bm x)$ that could be used to make a prediction of $y$ at any $\bm x$ within the range of the training points. 

Artificial NNs provide one way of building such a model. While there are many flavours of NNs, consider a simple NN with one hidden layer illustrated by Figure 1. 
Figure 1 is a schematic depiction of the following mathematical expression: 
\begin{eqnarray}
y(\bm x) = b + \sum_{j = 1}^{\cal H} v_j h_j(\bm x)
\label{NN}
\end{eqnarray}
with 
\begin{eqnarray}
h_j(\bm x) = \tanh \left (a_j + \sum_{i=1}^N x_i u_{ij} \right)
\label{tanh}
\end{eqnarray}
and $\cal H$ representing the number of hidden units (unfilled circles). Each hidden unit provides a non-linear transformation (\ref{tanh}) of the variables and the wedges depict the fitting parameters $b, v_j, a_j, u_{ij}$ found by training the NN.  
The tanh functions in Eq. (\ref{tanh}) ensure that a NN can describe a general, non-linear relationship between the original variables $\bm x$ and $y$.  
There are several different non-linear functions commonly used for NN fits (another popular choice is a sigmoid function). The function tanh is used here as a representative example. 
We denote the parameters of the model collectively by $\bm \theta = \left [b, v_j, a_j, u_{ij} \right ]$. 

\begin{figure}[t]
	\includegraphics[width=0.5\columnwidth]{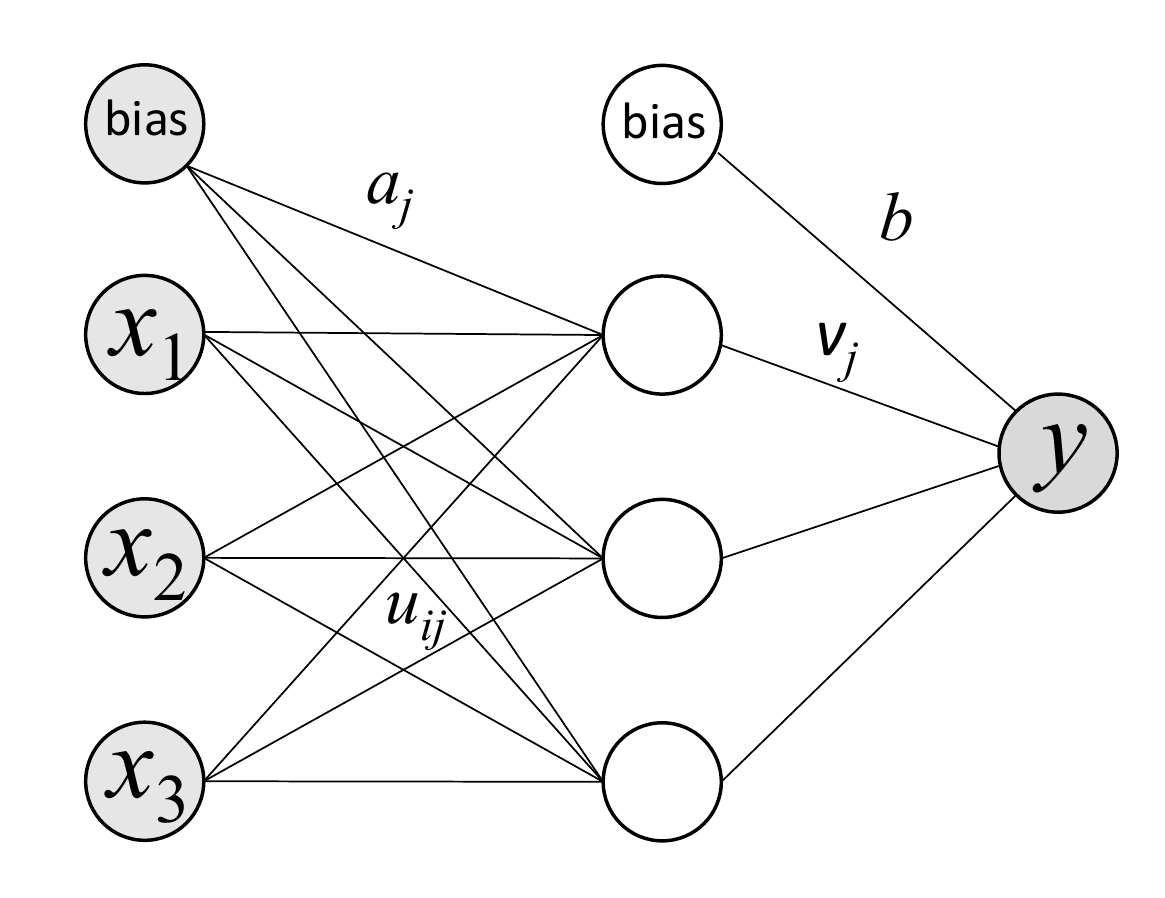}
	\caption{Schematic depiction of a feed-forward multi-layer perceptron neural network with one hidden layer represented by the unfilled circles. 
}
\end{figure}

\begin{figure}[H]
\centering
	\includegraphics[width=0.4\columnwidth]{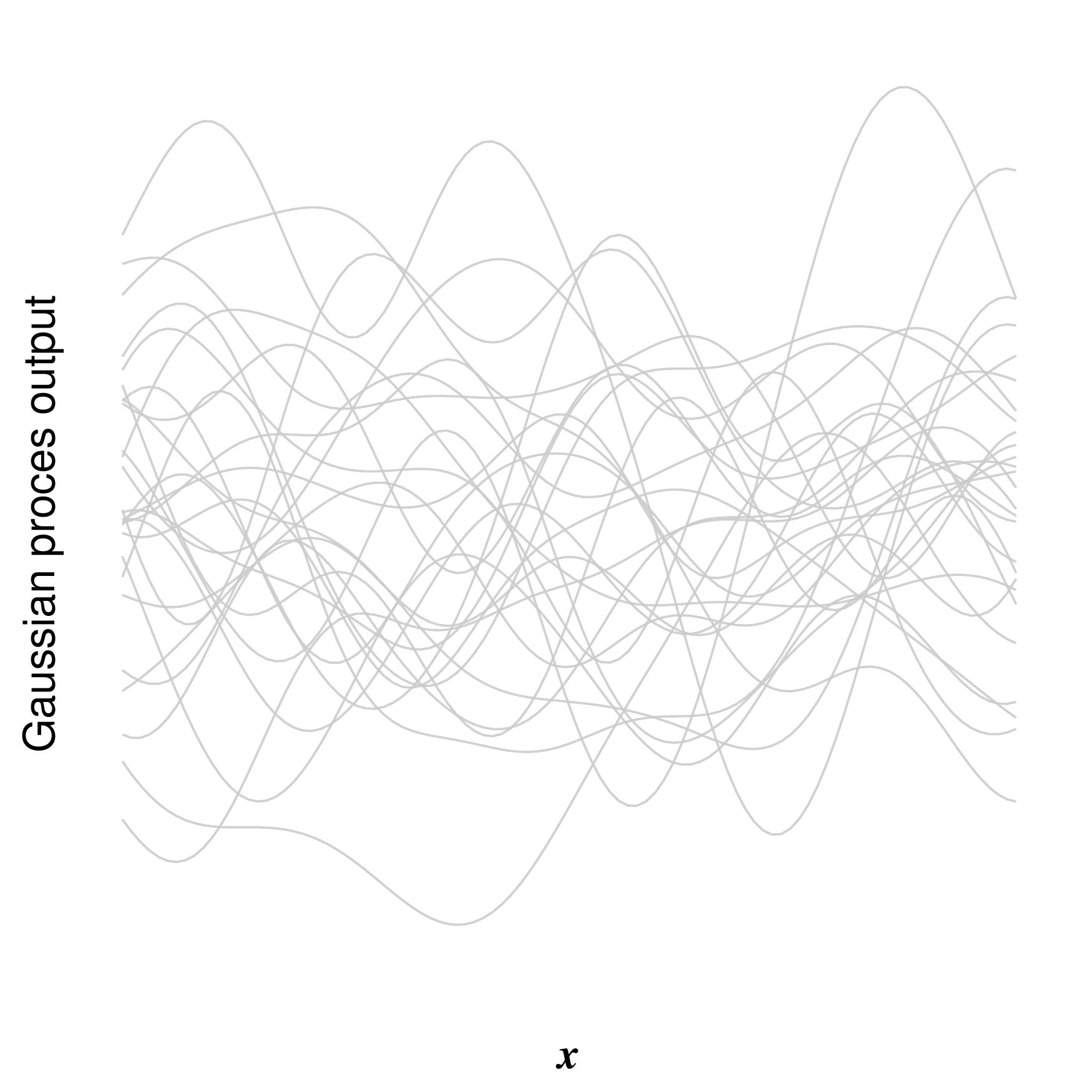}
		\includegraphics[width=0.4\columnwidth]{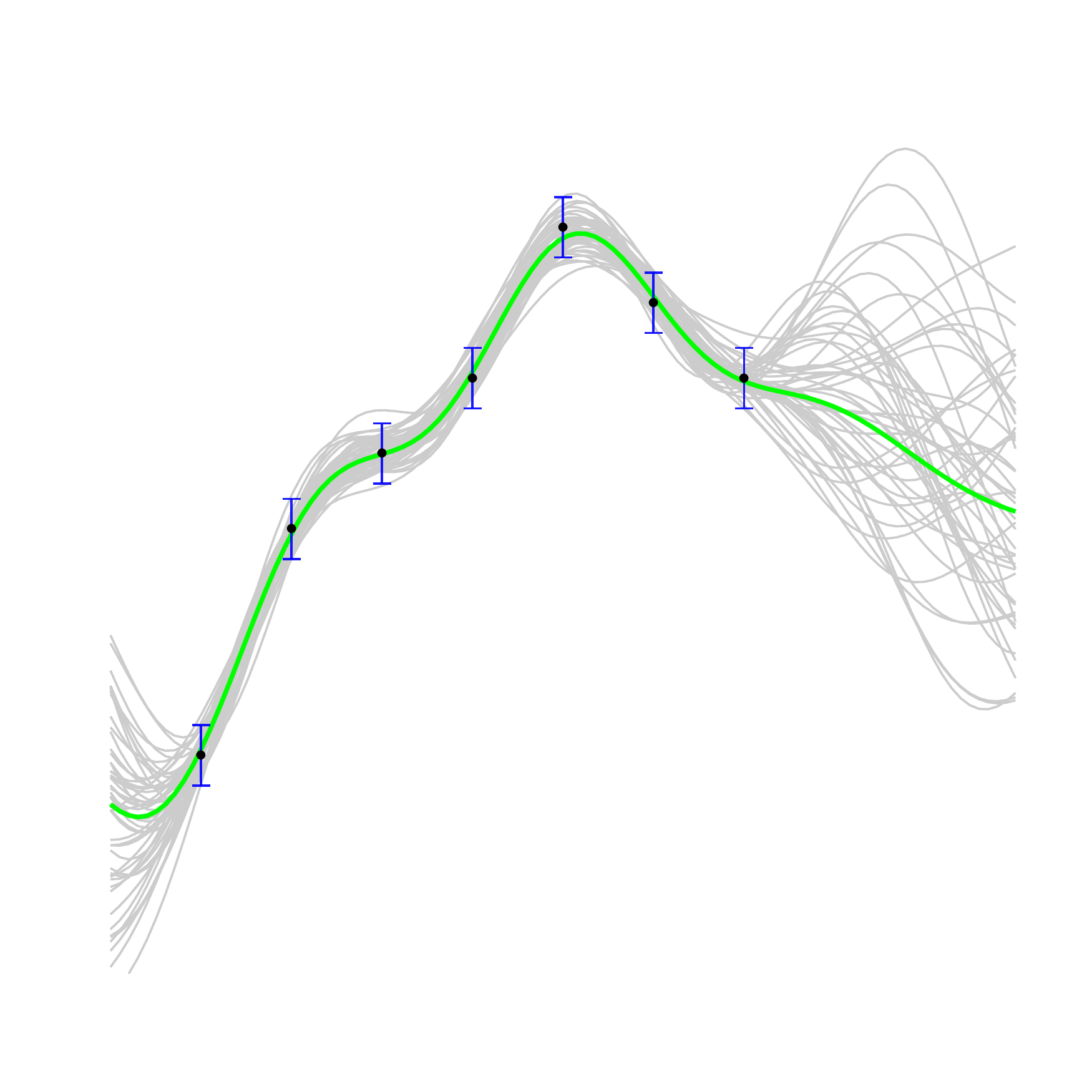} \\
		\includegraphics[width=0.4\columnwidth]{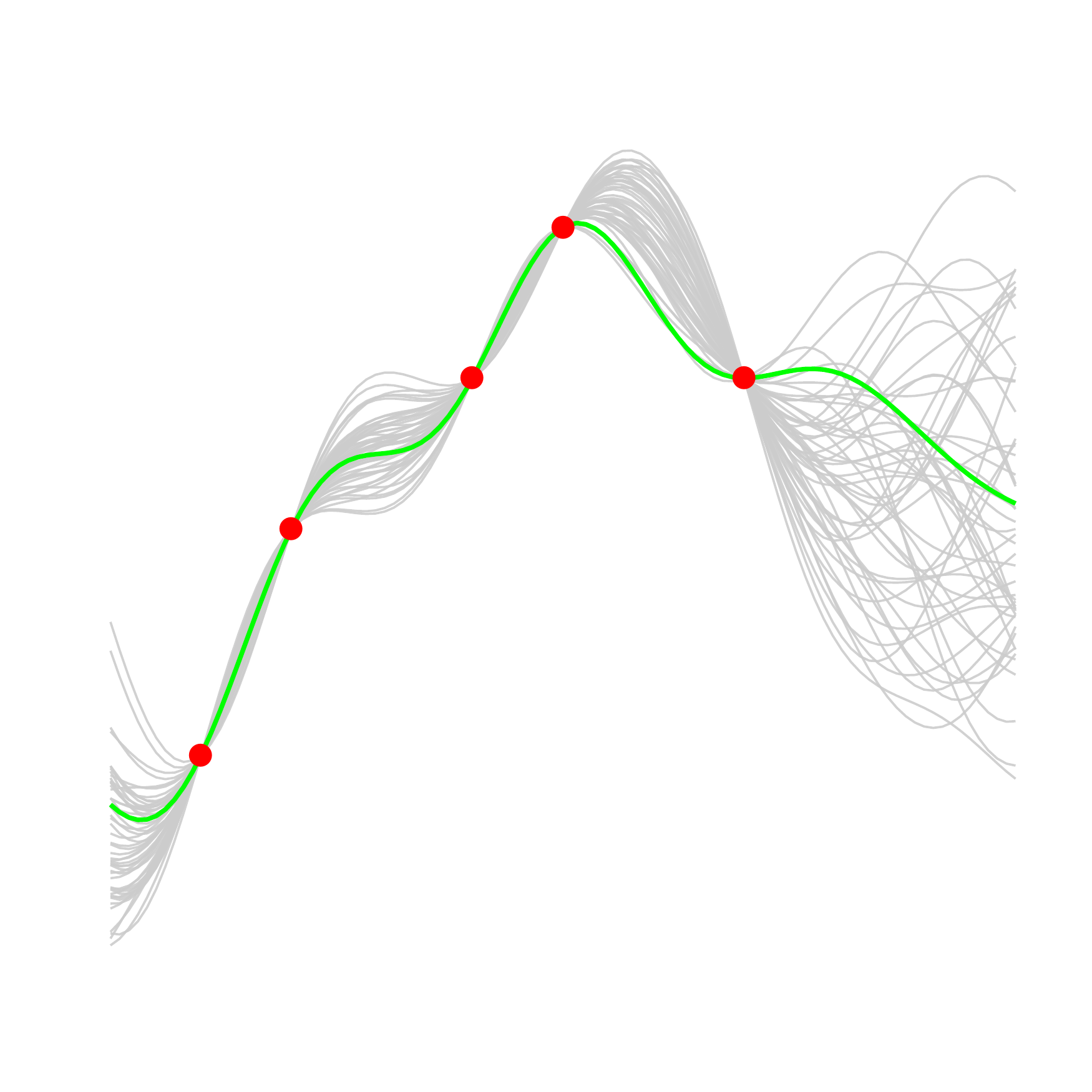}
		\includegraphics[width=0.4\columnwidth]{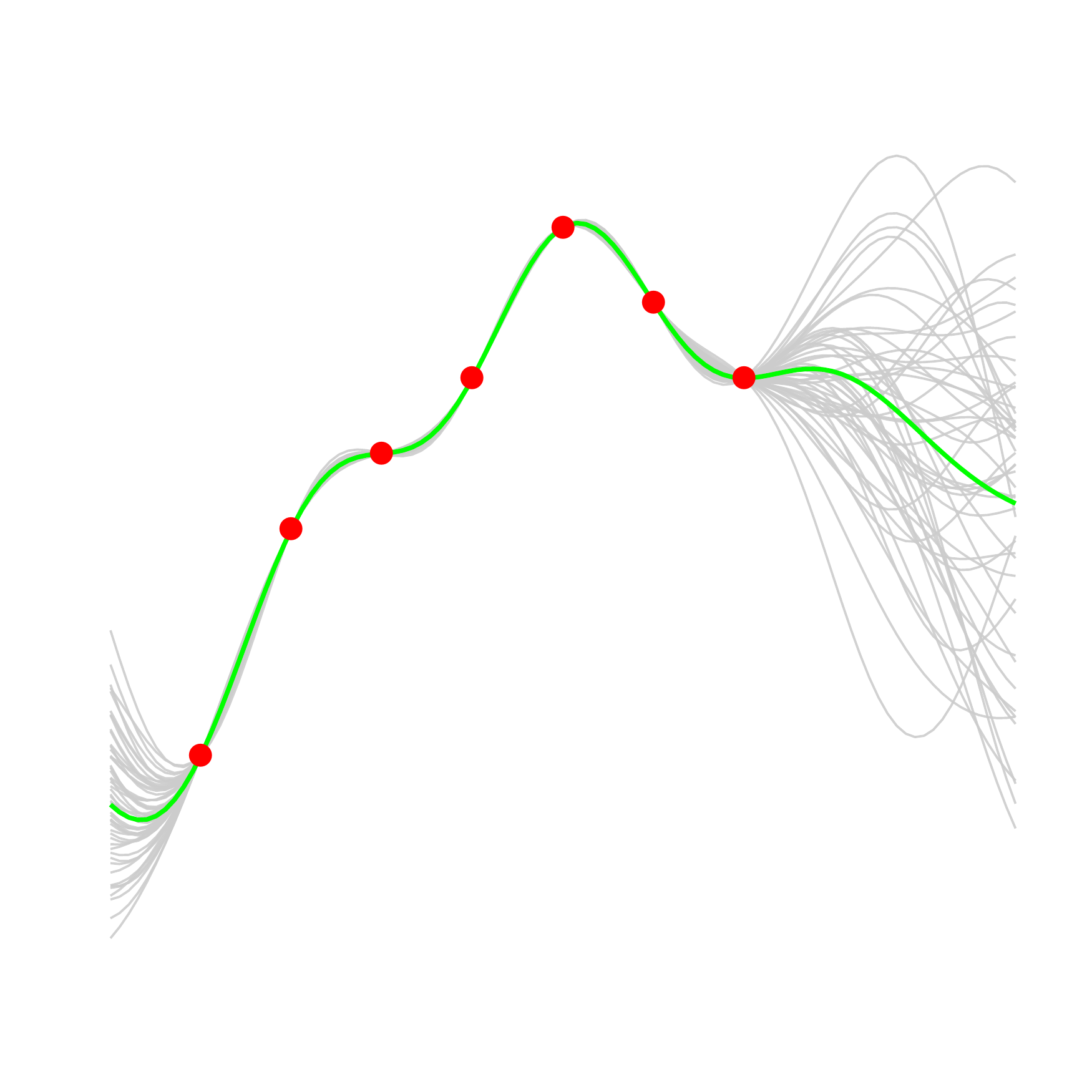}
	\caption{Illustration of Bayesian machine learning. The prediction model is built with Gaussian Processes. A GP can be viewed as a distribution of random functions normally distributed around some mean function. The {\bf top left panel} shows the Gaussian process prior, i.e. a collection of random functions normally distributed around zero with some variance chosen at will. This GP must be modified (conditioned) using the observations (training data). The {\bf bottom left panel} shows the Gaussian process (grey curves) conditioned by the observations (red symbols). The mean of such GP is given by Eq. (\ref{GP-mean}) yielding the green curve. The distribution of the grey curves illustrates the Bayesian uncertainty of the prediction between the training points.  The {\bf bottom right panel} shows how this uncertainty is reduced when more training points (symbols) are added. The {\bf upper right panel} shows the GP trained by observations with some inherent uncertainty (noise) illustrated by the blue bars. The plots are obtained with a R program adopted from Ref. \cite{gp-book}. 
}
\label{figure:BML}
\end{figure}

When the parameters $b, v_j, a_j, u_{ij}$ are fixed, we have a fixed model of $y(\bm x)$. However, one can also make a prediction of $y(\bm x)$ using a Bayesian approach. 
Within a Bayesian approach, the parameters $\bm \theta$ are not fixed. Rather, the parameters $\bm \theta$ are treated as random variables that have some distribution. 
One starts by defining a guess probability distribution $P(\bm \theta)$ of the model parameters, known as the prior.  This distribution reflects our belief about the model. It is defined without any training data and is simply based on reasonable assumptions about the parameters of the model. In principle, one can draw values of the parameters $\bm \theta$ from $P(\bm \theta)$, evaluate the model $y(\bm x, \bm \theta)$ at the locations of the training points and compare the results with the training data $\bm y$. 
This will define the likelihood for the model $P(\bm y |\bm \theta)$, i.e. the probability density of observing the correct values $\bm y = [y_1, y_2, ..., y_n]^\top$ {\it given the model parameters $\bm \theta$}. The goal is, however, to find $P(\bm \theta | \bm y)$, i.e. the probability distribution of the model parameters {\it given the training data}. $P(\bm \theta | \bm y)$ is called the posterior distribution. The two distributions are related by the Bayes' theorem: 
\begin{eqnarray}
P(\bm \theta | \bm y) = P(\bm y | \bm \theta) \frac{P(\bm \theta)}{P(\bm y)},
\end{eqnarray}
where 
\begin{eqnarray}
P(\bm y) = \int_{\bm \theta} P(\bm y | \bm  \theta) P(\bm \theta) {\rm d} \bm \theta
\end{eqnarray}
is the {\it marginal likelihood} for the model.  The word `marginal' means that the parameters have been `marginalized' (i.e. integrated over). 
The ultimate goal is to make prediction of the unknown value $y_\ast$ at $\bm x_\ast$, given the known values $\bm y$. This prediction can be made by integrating over the model parameters, to yield
\begin{eqnarray}
P( y_\ast | \bm y) = \int_{\bm \theta}  P( y_\ast | \bm \theta) P(\bm \theta | \bm y)  {\rm d} \bm \theta.
\end{eqnarray}
As we can see, the Bayesian approach produces a conditional probability distribution for $y_\ast$. One can use, for example, the mean of this distribution as the prediction value. 
Changing the training points $\bm y$ (for example, by adding more points to the set $\bm y$) changes the distribution $P(\bm \theta | \bm y)$ and hence the predictive distribution  
$P( y_\ast | \bm y)$. The Bayesian approach to building a model of an unknown function is illustrated in Figure \ref{figure:BML}.

Promoting the parameters $b, v_j, a_j, u_{ij}$ to random variables makes Eq. (\ref{NN}) a Bayesian Neural Network. Eq. (\ref{NN}) is no longer a single value for a given $\bm x$ but a distribution. 
Using the prior distributions for each of these parameters gives the prior distribution for $y(\bm x)$. Since the parameters of the model are variables, the model has to be written explicitly as $y(\bm x, \bm \theta)$. We should now choose the prior distributions for the model parameters. If it is possible to construct an accurate NN with fixed values of the parameters $b, v_j, a_j, u_{ij}$, it is reasonable to assume Gaussian distributions for each of $b, v_j, a_j, u_{ij}$.

If Gaussian priors are chosen, Eq. (\ref{NN}) becomes a sum of a Gaussian (the first term) and a sum over $\tanh(\gamma_i)$ multiplied by another Gaussian. If the number of terms in the sum (i.e. the number of hidden units of the NN) is small, we cannot draw any conclusions about the analytical form of the sum. However, if the number of hidden units goes to infinity ${\cal H} \rightarrow \infty$, one can use the central limit theorem to conclude that the second term in Eq. (\ref{NN}) is also a Gaussian \cite{BML}. 
Thus, if each of the parameters in $\bm \theta$ is Gaussian-distributed {\it and} the number of hidden units goes to infinity ${\cal H} \rightarrow \infty$, 
$y(\bm x)$ becomes Gaussian-distributed. In other words, the repeated evaluation of the model $y(\bm x)$ with parameters $b, v_j, a_j, u_{ij}$ randomly drawn from their distributions produces a Gaussian distribution. 
Since $y(\bm x)$ is a smooth function of $\bm x$, Eq. (\ref{NN}) thus becomes a Gaussian process.

There is an excellent book on Gaussian Processes for ML \cite{gp-book}. 
Gaussian processes are much easier to work with than Bayesian NNs because a GP is entirely determined by its mean function and covariance $ k(\bm x, \bm x') = {\rm Cov}(\bm x, \bm x')$, which describes the relationship between the Gaussian distributions $y(\bm x)$ and $y(\bm x')$ at two different points of the $N$-dimensional variable space. Also, because the prior distribution 
$P(\bm \theta)$ is Gaussian and the likelihood can be written as a product of Gaussians, the posterior and the predictive distributions are also Gaussian. One might argue that GPs make Bayesian ML practical. They certainly make Bayesian optimization (described in the following section) easy to implement. 

To see how GPs work in practice (and get a gist of the main idea behind kernel methods in ML), 
let us treat $h_j$ in Eq. (\ref{NN}) as new variables and introduce a function $\phi(\bm x)$ that maps an $N$-dimensional vector $\bm x$ onto the $h_j$ functions, which live in an $\cal H$-dimensional space. When one trains a NN as in the example above, one essentially is looking to determine this map.
However, this map does not need to be specified explicitly. Instead, one can formulate the learning algorithms using `kernels', as explained below.   
The $\cal H$-dimensional  space is called the {\it feature space}. As we can see from Eq. (\ref{NN}), the relationship between $y$ and $h_j$ is linear so $y$ is a linear function of $\cal H$ variables in the feature space. If we had the exact map of $\bm x$ onto the feature space, we could interpolate (and extrapolate) the linear function $y$ in the feature space, then map the result back to our original $N$-dimensional space to make accurate predictions of $y(\bm x)$ at any $\bm x$. 

Let us re-write Eq. (\ref{NN}) in matrix form 
\begin{eqnarray}
y(\bm x) = \bm \phi^{\top} \bm w
\label{linear-model}
\end{eqnarray}
where $\bm \phi$ is an (${\cal H} + 1$)-dimensional vector of $h_j$ functions and an additional bias element (whose value is always one), and $\bm w$ is a vector of $\cal H$ weights $v_j$ and $b$. 
We write the Gaussian prior distribution of $\bm w$ as
\begin{eqnarray}
P(\bm w) \propto \exp{(- {\bm w^{\top} \bm \Sigma^{-1} \bm w})} 
\label{model-prior}
\end{eqnarray}
where $\bm \Sigma$ is the covariance matrix.  


To obtain the expression for the likelihood, it is assumed that the observations (training points) have some random noise that can be modelled by a Gaussian distribution with variance $\sigma$ \cite{gp-book}. 
If the training data are noiseless, one can use the $\sigma \rightarrow 0$ limit of the final equations. 
 Consider a specific training point that has value $y_i$ at the position $\bm x_i$. Since the model deviates from the observation by random noise with variance $\sigma$, the probability density of the observation $y_i$ is 
\begin{eqnarray}
P(y_i| \bm w) = \frac{1}{\sqrt{2 \pi} \sigma} \exp{\left \{ - \frac{( y_i - \bm \phi^{\top}(\bm x_i) \bm w )^2}{2 \sigma^2} \right \}}.
\end{eqnarray}
 Assuming that the different training points are independent, we can then write the likelihood function as a product
\begin{eqnarray}
P(\bm y | \bm w) = \prod_{i = 1}^n P(y_i | \bm w) = \frac{1}{(2 \pi \sigma^2)^{n/2}}   \exp{\left \{ - \frac{| \bm y - \bm \Phi^{\top} \bm w |^2}{2 \sigma^2} \right \}},
\end{eqnarray}
where $\bm \Phi$ is a matrix with $n$ columns of $\bm \phi$ corresponding to the different training points and ${\cal H} + 1$ rows. 

Using the Bayes' theorem, we can now write the posterior distribution as 
\begin{eqnarray}
P(\bm w | \bm y) \propto P(\bm y | \bm w) P(\bm w) \propto \exp{(- {\bm w^{\top}\bm  \Sigma^{-1} \bm w})}  \times \exp{\left \{ - \frac{( \bm y - \bm \Phi^{\top} \bm w )^\top 
( \bm y - \bm \Phi^{\top} \bm w )}{2 \sigma^2} \right \}}
\\
= \propto \exp{ \left \{
(\bm w - \bar {\bm w})^\top
{\bm A}^{-1}
(\bm w - \bar {\bm w})
 \right \}    }
\end{eqnarray}
where $\bm A = \sigma^{-2} \bm \Phi \bm \Phi^\top + \bm \Sigma^{-1}$, $\bar{ \bm w} = \sigma^{-2}(\sigma^{-2} \bm \Phi \bm \Phi^\top + {\bm \Sigma}^{-1} )^{-1} \bm \Phi \bm y$ and all $\bm w$-independent terms have been omitted. This shows that the posterior distribution is a multivariate Gaussian with mean $\bar{\bm w}$ and covariance matrix $\bm A$. See Ref. \cite{gp-book} for a detailed derivation. 


In order to obtain the predictive distribution, we must multiply the posterior by $P(y_\ast | \bm w)$ and integrate over $\bm w$. This yields a Gaussian distribution with the mean \cite{gp-book}
\begin{eqnarray}
\mu_\ast =  {\bm \phi}^{\top}_{\ast} \bm \Sigma \bm \Phi (\bm K + \sigma^2 \bm I)^{-1} \bm y
\label{mean}
\end{eqnarray}
and variance
\begin{eqnarray}
\sigma_\ast = {\bm \phi}^{\top}_{\ast} \bm \Sigma \bm \phi_\ast - {\bm \phi}^{\top}_{\ast} \bm \Sigma \bm \Phi (\bm K + \sigma^2 \bm I)^{-1} \bm \Phi^\top \bm \Sigma \bm \phi_\ast,
\label{variance}
\end{eqnarray}
where $\bm \phi_\ast = \bm \phi(\bm x_\ast)$ and $\bm K = \bm \Phi^\top \bm \Sigma \bm \Phi$. 

These equations are useless because we do not know the maps $\phi(\bm x)$. However, one can see that all terms in Eqs. (\ref{mean}) and (\ref{variance}) include products such as $\bm \Phi^\top \bm \Sigma \bm \Phi$ or ${\bm \phi}^{\top}_{\ast} \bm \Sigma \bm \Phi$. Following Ref. \cite{gp-book}, one can define the matrix $\bm D = \bm \Sigma^{1/2}$ such that $\bm \Sigma = \bm D^\top \bm D $ and the $({\cal H} +1)$-dimensional vectors 
\begin{eqnarray}
\bm \psi(\bm x) = \bm D \bm \phi(\bm x). 
\end{eqnarray}
One can further denote the scalar products of these vectors by 
\begin{eqnarray}
 k(\bm x, \bm x') = \bm \psi(\bm x)^\top \bm \psi(\bm x'). 
\end{eqnarray}
With these definitions and notation, the vector ${\bm \phi}^{\top}_{\ast} \bm \Sigma \bm \Phi$ is a vector of scalar products $k(\bm x_\ast, \bm x_i)$
and the matrix $ \bm K = \bm \Phi^\top \bm \Sigma \bm \Phi$ is a square $n \times n$ matrix of the scalar products $k(\bm x_i, \bm x_j)$, where  $\bm x_i$ and $\bm x_j$ correspond to the locations of the training points in the original $N$-dimensional space containing $n$ training points.

Thus, Eqs. (\ref{mean}) and (\ref{variance}) are completely defined by the variance of the noise $\sigma$, the values of the training points $\bm y$ and the scalar products $k(\bm x, \bm x')$. 
To emphasize this, we re-write Eqs. (\ref{mean}) and (\ref{variance}) as 
\begin{eqnarray}
\mu_\ast = \bm k_\ast^\top (\bm K + \sigma^2 \bm I)^{-1} \bm y,
\label{GP-mean}
\end{eqnarray}
\begin{eqnarray}
\sigma_\ast = k(\bm x_\ast, \bm x_\ast)  - \bm k_\ast^\top  (\bm K + \sigma^2 \bm I)^{-1} \bm k_\ast
\label{GP-variance}
\end{eqnarray}
where $\bm k_\ast$ is a vector with $n$ entries $k(\bm x_\ast, \bm x_i)$.
The unknowns in these equations are the functions $k(\bm x, \bm x')$ and the variance $\sigma$. 
One typically specifies the mathematical form of the functions $k(\bm x, \bm x')$ based on some physical or mathematical assumptions (discussed below). Given a specific form of $k(\bm x, \bm x')$, one has a GP {\it model}. This model depends on the (at this point unknown) parameters of $k(\bm x, \bm x')$. 
These parameters are found by maximizing the marginal likelihood function, i.e. the integral  
\begin{eqnarray}
P(\bm y) = \int_{\bm w} P(\bm w| \bm y) P(\bm w) d \bm w.
\label{marginal-likelihood}
\end{eqnarray}
The larger the value of this integral, the better the model. 

The functions $k(\bm x, \bm x')$ map two inputs $\bm x$ and $\bm x'$ onto a real number (representing a scalar product of vectors in the feature space). Such mapping functions are called `kernels' \cite{gp-book}.  
The functional dependence of $k(\bm x, \bm x')$ on $\bm x$ and $\bm x'$ is called the kernel function. 
It is useful to examine the relation of these kernels to covariance of the GP (\ref{linear-model}). In general, for two real random variables  $\alpha = y(\bm x)$ and $\beta = y(\bm x')$, the covariance is defined as 
\begin{eqnarray}
{\rm Cov}(\alpha, \beta) =  \langle \left ( \alpha - \langle \alpha \rangle \right ) \left ( \beta - \langle \beta \rangle \right ) \rangle,
\end{eqnarray}
where $\langle \cdot \rangle$ denotes the expected value.
If a GP is based on the Gaussian model prior (\ref{model-prior}), this expected value is 
\begin{eqnarray}
{\rm Cov}(\bm x, \bm x')  =  \langle \bm \phi^{\top}(\bm x) \bm w \bm w^\top \phi (\bm x') \rangle = \bm \phi^{\top}(\bm x) \langle \bm w  \bm w^\top \rangle \bm \phi (\bm x') = 
\bm \phi^{\top}(\bm x) \bm \Sigma \bm \phi (\bm x')  = k(\bm x, \bm x')
\label{covariance-gp}
\end{eqnarray}
Thus, the covariance of such GP is the kernel $k(\bm x, \bm x')$. 

This is important because the dependence of the kernels on $\bm x$ and $\bm x'$ is generally unknown and Eq. (\ref{covariance-gp}) can inform the design of the kernel functions. In particular, if the training data are not periodic (and do not have long-range correlations), it is reasonable to assume that the covariance of the GP (\ref{linear-model}) decays with the distance $|\bm x - \bm x'|$. Therefore, in order to train a GP model, one  begins with assuming some mathematical function for $k(\bm x, \bm x')$ that decays with $|\bm x - \bm x'|$. The following examples are some of the most commonly used functions for approximating kernels used for GP regression: 
\begin{eqnarray}
k_{\rm LIN}({\bm x}, {{\bm x}'})  =  {\bm x}^\top {{\bm x}'}
\label{eqn:k_LIN}
\end{eqnarray}
\begin{eqnarray}
k_{\rm RBF}({\bm x}, {{\bm x}'}) = \exp \left(-\frac{1}{2}r^2({\bm x}, {{\bm x}'})\right)
\label{eqn:k_RBF}
\end{eqnarray}
\begin{eqnarray}
k_{\rm MAT}({\bm x}, {{\bm x}'})  = \left( 1 + \sqrt{5}r({\bm x},{{\bm x}'}) +  \frac{5}{3}r^2({\bm x}, {{\bm x}'})\right )
\nonumber
\\
\times \exp\left ( -\sqrt{5}r({\bm x}, {{\bm x}'})\right )~~~~
\label{eqn:k_MAT}
\end{eqnarray}
\begin{eqnarray}
k_{\rm RQ}({\bm x}, {{\bm x}'})  = \left ( 1 + \frac{|{\bm x}- {{\bm x}'}|^2}{2\alpha\ell^2} \right )^{-\alpha}
\label{eqn:k_RQ}
\end{eqnarray}
where $r^2({\bm x}, {{\bm x}'}) = ({\bm x}- {{\bm x}'})^\top \times {\bm M} \times ({\bm x}-{{\bm x}'})$ and ${\bm M}$ is a diagonal matrix with different length-scales $\ell_d$ for each dimension of ${\bm x}$. 
The labels of the kernels stand for `linear', `radial basis function', `Mat\'ern'  and `rational quadratic' kernels. 
These functions are parametrized by the length-scale parameters $\ell_d$, $\ell$ and $\alpha$. 
We denote these free parameters collectively by $\bm \gamma$.  
To train a GP model, one varies these parameters in order to maximize the marginal likelihood (\ref{marginal-likelihood}). In practice, it is easier to maximize the logarithm of the marginal likelihood. 
Because the prior of the GP and the likelihood function are Gaussian distributions, one can evaluate the logarithm of the integral (\ref{marginal-likelihood}) to have the following form \cite{gp-book}: 
\begin{eqnarray}
\log P({\bm y} | \bm \gamma) = -\frac{1}{2}{\bm y}^\top \left ( {\bm K} + \sigma^2 \bm I \right)^{-1}{\bm y} - \frac{1}{2}\log |\bm K + \sigma^2 \bm I| - \frac{n}{2} \log 2\pi. 
\label{log-likelihood-explicit}
\end{eqnarray} 

To summarize, a GP prediction of the property $y$ at a point $\bm x$ of an $N$-dimensional space begins with $n$ known values of $y$
collectively represented by vector $\bm y$. Given these values, a GP model is built by (i) assuming a particular mathematical form for the kernel function $k(\bm x, \bm x')$ parametrized by some unknown coefficients;  (ii) optimizing the logarithm of the marginal likelihood function (\ref{log-likelihood-explicit}) by iteratively computing the matrix $\bm K$ and its inverse; (iii) making a prediction of $y_\ast$
at $\bm x_\ast$ using Eq. (\ref{GP-mean}). The side benefit of this approach is Eq. (\ref{GP-variance}) that provides a Bayesian uncertainty of the prediction. Note that this uncertainty reflects the lack of complete knowledge of the function $y(\bm x)$ and it decreases with the number of training points in $\bm y$. The uncertainty of the observations themselves is described by the variance $\sigma$. If the training points are noiseless (as will often be the case in this article), Eqs. (\ref{GP-mean}), (\ref{GP-variance}) and (\ref{log-likelihood-explicit}) should be used with $\sigma$ set to zero. The choice of the kernel function affects the efficiency of the learning but accurate interpolation is possible with many different mathematical forms of the kernel function. We will discuss the effect of the choice of the kernel function in subsequent sections.

\subsection{Bayesian optimization}

\begin{figure}[t]
	\includegraphics[width=0.8\columnwidth]{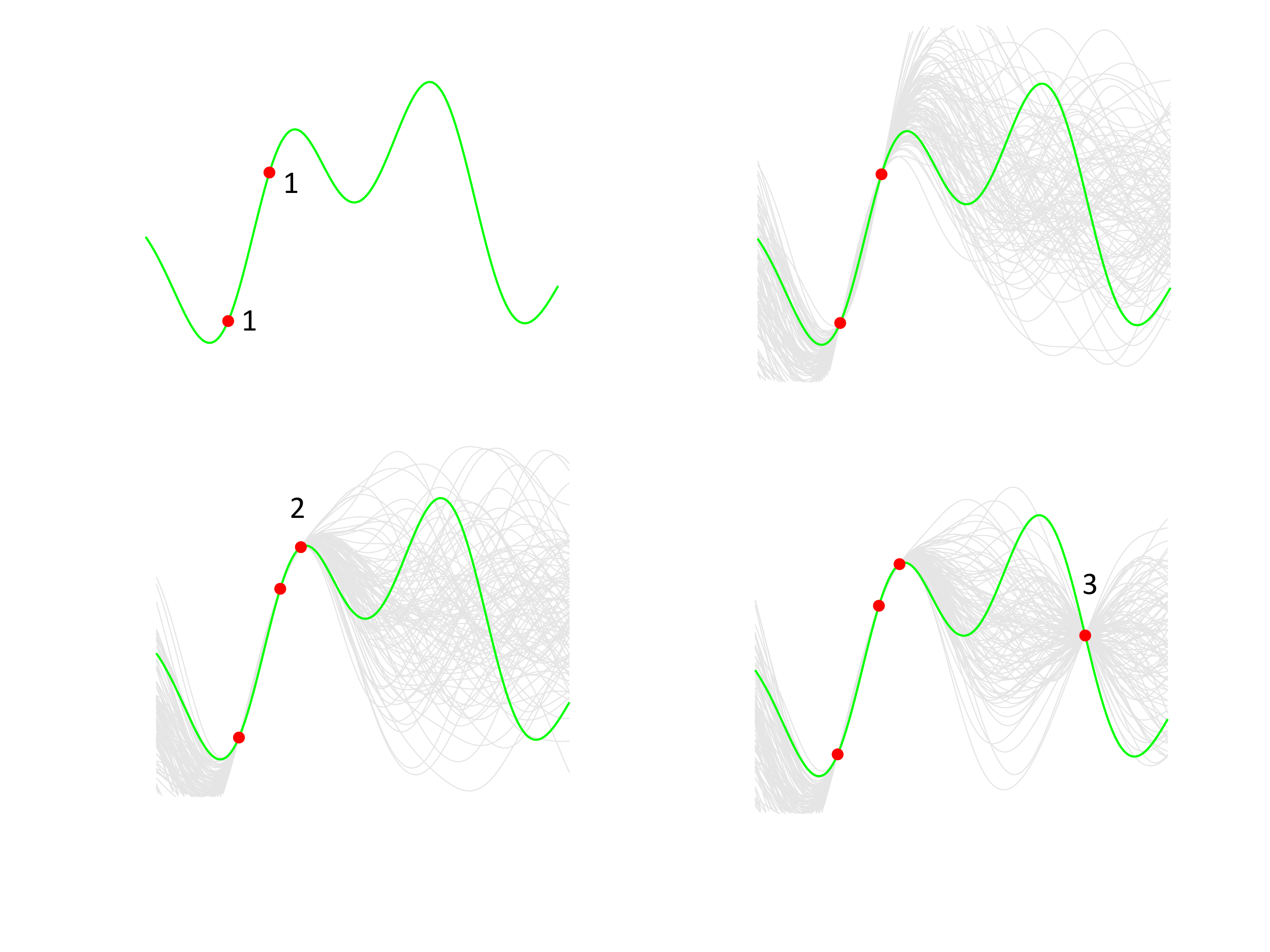}
	\caption{Bayesian optimization of a function (shown by solid green curve) by the iterative evaluation of the function. A new GP model of the function (grey curves) is constructed at each iteration and the subsequent evaluation of the function is informed by the mean and the uncertainty of the most recent GP. The symbols depict the results of the function evaluation and the numbers label the iteration order.  Each function evaluation reduces the uncertainty of the resulting GP in some part of the variable space, which forces the algorithm to evaluate the function elsewhere. This ensures that the optimization algorithm does not get trapped in local extrema. In the present example, the algorithm begins by evaluating the function at the points labeled `1'. The acquisition function (described in text) directs the subsequent evaluation to point `2', as this is where the mean of the GP and hence the expected value of the function are greater than the values at points `1'. The acquisition function then directs the subsequent evaluation to point `3', where the GP has a large uncertainty (and hence an unknown probability of the function to have an even greater value). 
}
\label{figure:BO}
\end{figure}

The approach described in the previous section can be used for regression and classification ML problems. It can also be used to optimize unknown functions that are difficult to evaluate. 
In applications advocated by the present work, the function $y(\bm x)$ often represents the solutions of the Schr\"{o}dinger equation that are neither known analytically nor easy to obtain numerically (see Section IV.A for details). Therefore, Bayesian optimization (BO) is particularly useful for applications in quantum molecular dynamics. 

The idea of BO can be described as follows. Consider an unknown, multidimensional function $y(\bm x)$. As mentioned, the relationship between $\bm x$ and $y$ does not need to be analytic. It is rather general. For example, $\bm x$ could be the parameters of a particular Hamiltonian and $y$ could be a particular eigenvalue of this Hamiltonian. The goal of BO is to find the global extremum of $y(\bm x)$ with as few evaluations of $y(\bm x)$ as possible and without computing the gradient of $y(\bm x)$. BO begins with a few random evaluations of $y(\bm x)$ at a few random points of the $N$-dimensional variable space. The results of the evaluation are used to obtain  Eqs. (\ref{GP-mean}) and (\ref{GP-variance}) giving an approximate model of $y(\bm x)$ and a Bayesian uncertainty of this model. Both of Eqs. (\ref{GP-mean}) and (\ref{GP-variance}) are then used to inform the subsequent evaluation of $y(\bm x)$. In the simplest possible formulation, this can be achieved by evaluating $y(\bm x)$, where the function $\alpha(\bm x) = \mu_\ast (\bm x) + \kappa \sigma_\ast (\bm x)$ has a maximum. 
The function $\alpha(\bm x)$ is called the `acquisition function'. 
Depending on the value of $\kappa$, the subsequent evaluation of the function $y(\bm x)$ is driven by $\mu_\ast$ (which forces the algorithm to find extrema of $y(\bm x)$ more accurately) or by $\sigma_\ast$ (which forces the algorithm to explore the $N$-dimensional space broadly). This is illustrated in Figure \ref{figure:BO}. 
 In the language of ML, the function $\alpha$ provides a balance between `exploitation' and `exploration'. 
More evaluations of the function $y(\bm x)$ in some part of the variable space reduce the uncertainty $\sigma_\ast$ of the GP process at this part of the space, which forces the algorithm to look elsewhere (where $\sigma_\ast$ is large) in this space. Depending on the application, one may choose to use a more complex acquisition function. 


BO offers several advantages over other optimization algorithms. First of all, it is very efficient (as will be demonstrated in a subsequent section). Second, it does not require the function gradient. Third, it does not rely on any properties of the function. The function does not have to be smooth or may even contain divergencies. The algorithm can be forced to ignore the divergencies and look for well-defined extrema. All decisions regarding function evaluations are based on GPs, whose mean and variance are necessarily smooth. This has made BO a powerful tool in ML. Perhaps, the most important application of BO in ML is training NNs. Traditionally, training a NN has been an art that requires an experienced user. BO can be used to automate the process and take the human element away from training a NN. 

\subsection{Extrapolation of physical results by generalization with machine learning}

It is uncommon to use ML for extrapolation of functions. 
ML is commonly used to make predictions of $y(\bm x)$ {\it within} the range of the training data. In the case of NNs, the prediction is in the form of a very complex fit of the observations. In the case of GPs, the prediction interpolates the observations. This statement immediately makes clear why the specific choice of the kernel function is not unique.  As illustrated in Figure (\ref{figure:BML}), if the training points are noiseless, the mean of the resulting GP must pass through  the training points. If the number of training points becomes large $n \rightarrow \infty$, a GP model with any (reasonable) kernel should represent accurately any function. As also illustrated by Figure \ref{figure:BML}, the GP becomes quite uncertain outside the range of the training points. 

  For applications in chemistry and physics, it would be very useful to develop ML methods capable of extrapolating physical results beyond the range of the training data. Such models could predict the physical properties in parts of the variable space that cannot be probed by rigorous theoretical calculations or experiments. Thus, such models could complement the common discovery tools. Predictions of new phase transitions is an example of an application of such models \cite{extrapolation-0}. 
  
  In the context of ML, it is more suitable to use the term `generalization' than `extrapolation'. `Generalization' refers to the ability of a ML model to describe previously unseen data. It is usually, though not always \cite{ml-extrapolation},  implied that these new, previously unseen, data come from the same distribution as the training data. In the present article, I will not attempt to make distinction between generalization and extrapolation. I use the term `extrapolation' to refer to predictions $y(\bm x^\ast)$ of a given physical property at a point $\bm x^\ast = \left [ x^\ast_1, ..., x^\ast_N \right ]$ in the $N$-dimensional space, if, at least, one of the variables $x^\ast_i \in \left [ x^\ast_1, ..., x^\ast_N \right ]$ is {\it outside} of the range covering the training data.

  Although extrapolation of solutions of physical equations with NNs has been attempted \cite{extrapolation-with-NNs-1,extrapolation-with-NNs-2}, it is a notoriously difficult problem, because of the complexity of the NN fit and the lack of information to constrain the behaviour of the NN outside the range of the training points. 
As explained in Section II.A, the use of kernels reduces the number of model parameters. 
In addition, the Bayesian approach described in Section II.A is designed to reject bad models and use good models with as little complexity as possible. Thus, the Bayesian approach should infer a linear dependence for a collection of points on a straight line, whereas a NN, if trained carelessly, might fit the same points with a sum of polynomials. As stated by R. N. Neal of the University of Toronto \cite{BML}, ``The Bayesian approach takes modelling seriously.''
One can thus argue that, in general, the Bayesian approach, and, particularly, kernel methods based on the Bayesian approach, are better suited for extrapolation of physical properties. 

This makes GPs interesting candidates for building ML models capable of extrapolation. 
In order to build GPs suitable for extrapolation, one should build models that are both more flexible and more physical. 
Improving the flexibility of the GP models can be achieved by making the kernel functions more complex. Improving the quality of the model can be achieved by maximizing the marginal likelihood. However, a simpler model producing the same outcome as a more complex model should always be preferred to avoid overfitting. It is, therefore, important to develop a metric that could quantify the quality of models with different kernel complexities. One such metric is the Bayesian information criterion (BIC) defined as \cite{bic}: 
\begin{eqnarray}
\text{BIC}({\cal M}) = \log P(\bm y |  {\cal M}) -\frac{1}{2}|{\cal M} |\log n,
\label{BIC-eq}
\end{eqnarray}
where $\cal M$ is the number of parameters in the kernel. The larger the BIC, the better the model. 


Using the BIC as a guiding quantity, one can build up the complexity of GP kernels to improve the predictive power of the models. Refs. \cite{extrapolation-1} and \cite{extrapolation-2} proposed to increase the complexity of the kernels by combining the simple kernels (\ref{eqn:k_LIN}) - (\ref{eqn:k_RQ}). It is generally impossible to train and try GP models with all possible combinations of even as few as four simple kernels. However, the complex models can be built using the following iterative algorithm, known as the `greedy' algorithm in reinforced learning.  
One begins by training a GP model with  {\it each} of the kernels (\ref{eqn:k_LIN}) - (\ref{eqn:k_RQ}) separately. 
The algorithm then selects the model with the highest value of the BIC. The kernel of this model  -- denoted $k_0$ --  is chosen as the base kernel. 
The base kernel $k_0(\cdot,\cdot)$  is then combined
   with each of the original kernels $k_i$ defined by Eqs. (\ref{eqn:k_LIN}) - (\ref{eqn:k_RQ}).
The kernels are combined as products $k_0(\cdot,\cdot)\;\times \;k_i(\cdot,\cdot)$ and additions $k_0(\cdot,\cdot)\;+ \;k_i(\cdot,\cdot)$.  
For each of the possible combinations, a new GP model is constructed and the BIC is computed. The kernel yielding the highest BIC is then used as a new base kernel $k_0$ and the algorithm is iterated.  As will be illustrated in a subsequent section, this approach leads to powerful GP models capable of predicting the physical properties of quantum systems outside the range of the training data. 
The question I would like to raise in this work is whether this tool can be of use in molecular dynamics research.

\section{Challenges in quantum molecular dynamics}

The objective of quantum dynamics calculations in theoretical chemistry is to solve the the Schr\"{o}dinger equation in order
to predict the outcome or understand the mechanisms of microscopic interactions of molecules with other molecules or molecules with electromagnetic fields.  
This is (almost) always done assuming the Born-Oppenheimer approximation leading to adiabatic potential surfaces of electronic energy. Even when dynamics is non-adiabatic, the problem is formulated in terms of potential energy surfaces, whether diabatic or adiabatic, and non-adiabatic couplings. The Hamiltonian of a system of $M$ atoms describing the nuclear dynamics is thus most generally written as 
\begin{eqnarray}
\hat H = \hat H(\left \{ \bm V(\bm r) \right \}),
\label{Hamiltonian}
\end{eqnarray}
where $\bm V (\bm r)$ is a set of $N$-dimensional potential energy surfaces (PES) and relevant $N$-dimensional couplings with $N = 3 M - 6$ (for non-linear polyatomic systems). If the dynamics is fully adiabatic, the Hamiltonian is parametrized by a single $N$-dimensional PES $V(\bm r)$. The variables $\bm r$ are the $N$ internal coordinates of the $M$-atom complex. For simplicity, we assume hereafter that only one PES is required to describe the dynamics, unless otherwise stated. 

The traditional approach always involves three steps: (1) computing the electronic energy of the system as a function of $\bm r$; (2) fitting the results of this computation with some $N$-dimensional analytical function; (3) solving the Schr\"{o}dinger equation to compute the desired observables. The Schr\"{o}dinger equation is solved using either a 
time-dependent or time-independent approach. Whether the Hamiltonian is time-dependent or time-independent, the problem can always be formulated as an eigenvalue problem by expanding the eigenfunctions of the  Hamiltonian (\ref{Hamiltonian}) is some basis. For example, the time-dependence of the eigenstates of periodic Hamiltonians can be described using the Floquet basis \cite{molecules-in-fields} and the radial dependence of the eigenstates of any Hamiltonian can be described using a discrete variable representation basis \cite{DVR}. 

A more practical approach is to represent the dependence of the Hamiltonian eigenstates on some variables by a basis set expansion, while leaving the dependence on the remaining variables explicit \cite{zhang-book}.  For example, for time dependent problems it is often most convenient to treat the time dependence explicitly. For time-independent reactive scattering problems, it is often convenient to formulate the problem using hyperspherical coordinates, with the hyperradius treated as an explicit variable.  
This converts the Schr\"{o}dinger equation into a system of coupled differential equations. If only one of the variables  -- variable $r$ -- is treated explicitly and the dependence on the other variables is represented by a basis set expansion, this system of differential equations can 
most generally be written as: 
\begin{eqnarray}
\left [ \hat {\cal D}_{r} \bm I + {\bm U} (V) \right ] \bm \Psi(r) = 0 
\label{the-differential-equation}
\end{eqnarray}
where $\hat {\cal D}_r$ is a differential operator acting on functions of $r$, 
$ \bm \Psi(r)$ is a vector of $N_b$
basis set expansion coefficients, each depending on $r$, 
${\bm U}$ is an $N_b \times N_b$ Hermitian matrix that depends on $V$, and  
$\bm I$ is the $N_b \times N_b$ identity matrix.  Given the PES $V(\bm r)$, Eq. (\ref{the-differential-equation}) can be solved numerically subject to appropriate boundary conditions in order to compute observables such as the bound state energies for spectroscopy applications, the $S$-matrix describing the probabilities of molecular collision outcomes or the chemical reaction rates. 
These calculations, however, meet with a lot of challenges. 

\subsection{Specific challenges}

There are two major problems with Eq. (\ref{the-differential-equation}). First, the numerical difficulty of solving this system of equations can scale up as quickly as ${\cal O}(N_b^3)$. 
The number of basis states $N_b$ required increases quickly with the number of degrees of freedom $N$. 
This makes the computation of the quantum observables for polyatomic systems very difficult. 
At present, numerically exact integration of the Schr\"{o}dinger equation is only possible for 2, 3 and 4-atom systems in the absence of external fields. 
The presence of fields breaks the isotropy of space, which results in a dramatic increase of the number of basis function $N_b$ required for the accurate representation of the eigenstates of the Hamiltonian \cite{molecules-in-fields}. The reactive scattering problem in the presence of external fields has only been solved rigorously for three-atom systems \cite{timur-1,timur-2}. 

The second problem is that the matrix $\bm U$ is parametrized by the $N$-dimensional PES $V(\bm r)$. This is a problem because quantum chemistry calculations always come with errors so $V(\bm r)$ is never exact. Therefore, even if one could obtain the numerically exact solutions of Eq. (\ref{the-differential-equation}), the results would still be affected by the uncertainty in $V$. 
Worst of all, it is often unknown how this uncertainty in PES affects the results of the quantum dynamics calculations. The dynamical results for different energies are affected by different parts of the PES to a varying degree of extent. This makes the predictions of quantum dynamics calculations unreliable, except where direct comparison with experiment is possible and the theoretical results are properly calibrated by the experimental data. 

Additional challenges include calculating the electronic energies for the potential energy surface and constructing the $N$-dimensional fit of $V(\bm r)$. While high-level quantum chemistry calculations are nowadays feasible for a large variety of molecular systems, including molecular radicals, it is not always clear how to sample the $N$-dimensional space by the electronic structure calculations. Consider, for example, a chemical reaction of two NaK molecules, of importance to ultracold molecule and ultracold chemistry experiments \cite{nak}. The PES required for the quantum dynamics computations of reaction probabilities in NaK - NaK collisions at low temperatures needs to be accurate at large molecule - molecule separations, 
represent accurately three- and four-body interactions, describe conical intersections, and be especially accurate along the minimum reaction path. The geometric features of this six-dimensional (6D) reaction complex are not known {\it a priori} and it is not clear how to sample this 6D space with quantum chemistry calculations. 

Computing accurate energies for the PES is, however, only half the problem. In order to be used in Eq. (\ref{the-differential-equation}), this surface must be represented by a proper fit or a 6D interpolating function, capable of extrapolation to large interatomic distances. Constructing such a fit is generally a major task that requires manual work. This task becomes significantly harder, as $N$ increases \cite{fitting-PES}, especially because the fit must properly account for all the geometrical features of the systems, including permutation symmetries \cite{fitting-PES-2}.  It is extremely important, especially for quantum dynamics calculations at low energies, to ensure that any such fit is free of artifacts and that the error of the fit is smaller than the uncertainty of the quantum chemistry calculations themselves. 

Yet another challenge in quantum molecular dynamics is related to esoteric dynamical features such as resonant scattering \cite{resonances,resonances2}. Resonances affect the dynamics of molecular interactions in a narrow range of collision energies \cite{resonances2} or a narrow range of applied field parameters \cite{timur-1}. This means that in order to identify all resonance features, Eq. (\ref{the-differential-equation}) must be solved on a very fine grid of total energies (for time-independent problems) and/or external field parameters. This makes the numerical integration of Eq. (\ref{the-differential-equation})  prohibitively difficult. In some cases, the dynamical features could be identified by solving a smaller set of equations obtained from Eq. (\ref{the-differential-equation}) by some decoupling approximations. However, in many cases, these approximations introduce additional sources of error and may affect the dynamical features of interest. 

With these challenges in mind, we identify the following major problems that, if solved, would transform the research field of quantum molecular dynamics:

\begin{itemize}
\item Reducing the number of coupled equations $N_b$ in Eq. (\ref{the-differential-equation}) without loss of accuracy of the resulting solutions. Of particular importance would be the development of methods that would reduce the scaling of the numerical complexity with the number of dimensions $N$, without introducing any approximations. Not all basis states in a particular basis set are equally important  for the accurate representation of quantum dynamics. A general  and easy-to-implement approach that could identify the `unnecessary' basis states would significantly reduce the CPU requirements for the numerical integration of Eq. (\ref{the-differential-equation}). 

\item Developing methods to evaluate the error bars of the dynamical results stemming from the uncertainties of the quantum chemistry calculations used to compute $V(\bm r)$. 
This is important in order to make absolute predictions of dynamical properties without calibration by experiment. 

\item Developing automated methods of constructing global $N$-dimensional PES $V(\bm r)$ without manual work and without loss of accuracy. 

\item Developing methods that could use approximate dynamical results (such as ones based on decoupling approximations) in order to predict accurate solutions of Eq. (\ref{the-differential-equation}). In the context of ML, this would amount to developing models that `learn' the error of the approximate dynamical calculations. 

\item Developing methods for solving the inverse scattering problem, i.e. constructing the PES describing interactions of molecules based on known experimental results of dynamical observables.

\end{itemize}
The purpose of this article is to show that ML can help solve these problems.

\section{Machine learning to address the challenges}

Solving the Schr\"{o}dinger equation is difficult and generally requires a lot of computational resources. Making a prediction with a ML model, once the model is trained, is fast and does not generally require significant computing power. The most obvious application of ML in quantum molecular dynamics is, therefore, to replace Eq. (\ref{the-differential-equation}) with a ML model that `simulates' the solutions of the Schr\"{o}dinger equation \cite{jie-prl}. The problem can be formulated as follows.

\subsection{ML models as simulators of Schr\"{o}dinger equation solutions}

Consider a specific observable such as a rate constant for a chemical reaction. Such rate constants can be computed by solving Eq. (\ref{the-differential-equation}) for a range of collision energies $E$ and a range of total angular momenta $J$ of the reactant molecules in order to compute the reaction probabilities and integrating the solutions over the Maxwell-Boltzmann distribution. If the reaction occurs in an external field, the computations must be repeated for different field parameters, such as the field amplitude, polarization and frequency, collectively denoted by $\bm f$.  The computations are performed using a PES often represented by an analytical fit with parameters $\bm a = \{ a_1, a_2, ...a_{k}, ... \}$. If one desires to explore the sensitivity of the resulting rate constants to the PES parameters or the field parameters, the computations must be repeated at different values of $\bm a$ and $\bm f$ and generally on a dense grid of collision energies and/or total angular momenta.  

Within a ML approach, the problem can be formulated as a model $y(\bm x)$, where $y$ represents the solution of the Schr\"{o}dinger equation (the reaction probabilitiy in our example), and $\bm x = \left [ E, J, \bm f, \bm a \right ]^\top$ is a vector of all parameters defining the Schr\"{o}dinger equation (\ref{the-differential-equation}). Instead of numerically solving Eq. (\ref{the-differential-equation}) for any desired combination of $\bm x$, one can perform a series of computations to determine the solutions of Eq. (\ref{the-differential-equation}) at random combinations of $\bm x = \left [ E, J, \bm f, \bm a \right ]^\top$. This will produce a set of values for the reaction probabilities $\bm y = \left [ y_1, y_2, ..., y_n \right ]^\top$. These values can be used as training data to build a ML model, as described in Section II. This model can then be used as a simulator to predict (with minimal numerical effort) the values of the reaction probabilities at any combination of  the parameters $\bm x = \left [ E, J, \bm f, \bm a \right ]^\top$. The rate constants can be calculated from these predictions, by simply integrating $y(\bm x)$ over the energy. The key advantage offered by kernel-based ML is that accurate simulators $y(\bm x)$ can be built with a small number of the Schr\"{o}dinger equation solutions.

It is important to note that the numerical effort of evaluating $y(\bm x)$ depends on the ML model used. If $y(\bm x)$ is represented by a NN, the prediction is the evaluation of an analytical function and is almost effortless. If $y(\bm x)$ is a Gaussian process, the prediction is in the form of a vector - vector product (\ref{GP-mean}), where the size of the vector is equal to the number of training points $n$, usually $\leq 5000$. The numerical cost of evaluating such models scales with the number of training points as ${\cal O}(n)$. 

Once the simulator $y(\bm x)$ is built, it can be used as a powerful tool to examine the mechanisms of microscopic molecular dynamics, obtain the dynamical results in a wide range of Hamiltonian parameters and obtain the error bars of the dynamical results, as explained in the following subsection.  
If one uses the Bayesian ML approach described in Section II, the accuracy of the simulator model must increase monotonously with the number of training points. 
For such simulators, the more solutions of the Schr\"{o}dinger equation are used to construct the models, the more accurate the predictions, as illustrated in Figure \ref{figure:BML}.

\begin{figure}[!ht]
\centering
	\includegraphics[width=0.6\columnwidth]{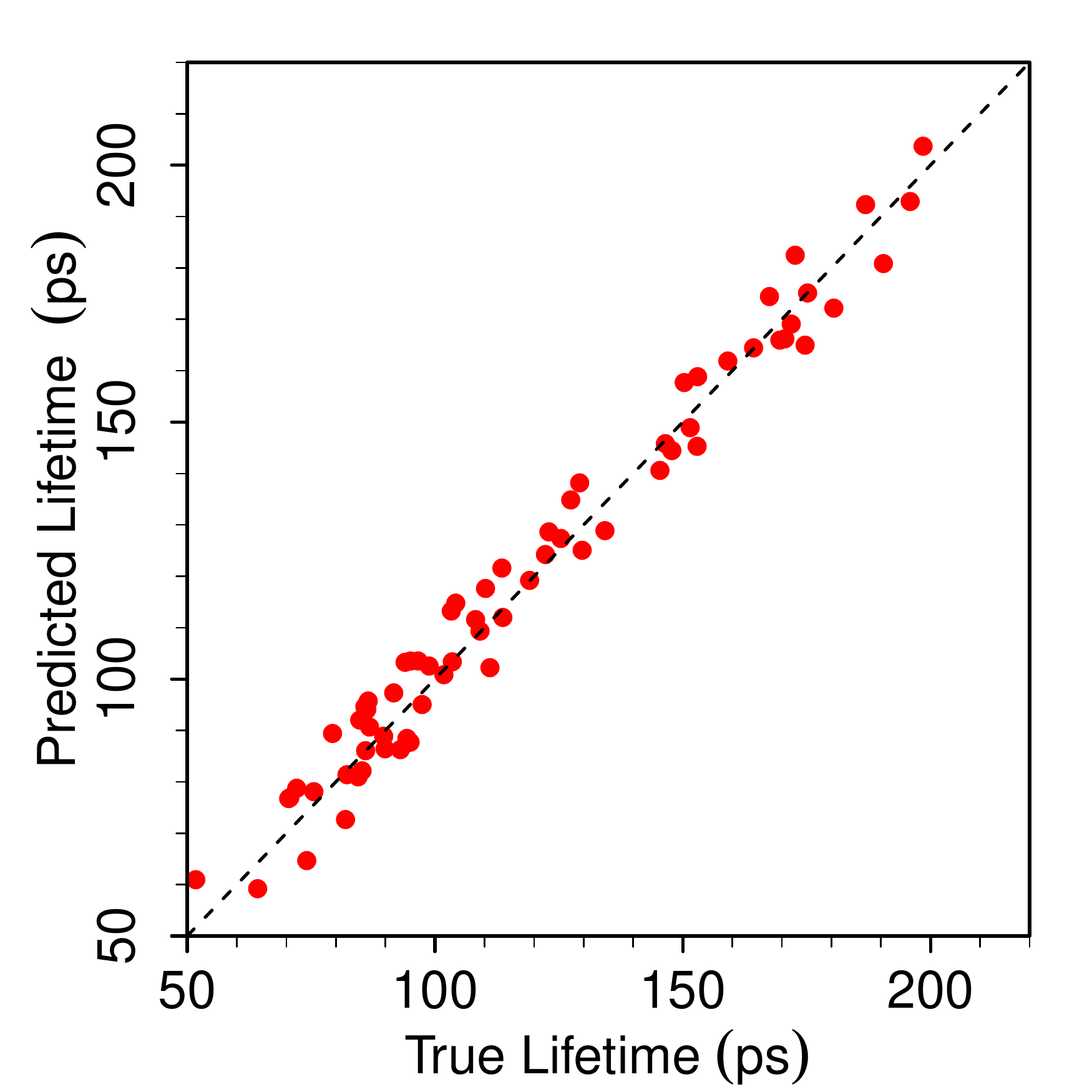}
	\caption{The comparison of the results of classical dynamics calculations with the predictions of the GP model trained by 200 dynamical calculations at 100 random combinations of 10 parameters specifying the Ar - C$_6$H$_6$ PES, the collision energy and the internal energy of the molecule. The results show the lifetime of the Ar - C$_6$H$_6$ collision complex. Figure adapted with permission from Ref. \cite{jie-prl}. 
}
\label{figure:ar-benzene}
\end{figure}

An example of a ML simulator of a dynamical property is illustrated in Figure \ref{figure:ar-benzene}. 
Figure \ref{figure:ar-benzene} considers the dependence of the collision lifetimes in Ar - C$_6$H$_6$ scattering on the PES parameters, the collision energy and the internal energy of the molecule.  The PES is expressed as a sum 
over terms describing the interaction of Ar with the C -- C and C -- H bond fragments \cite{ar-benzene}. There are 8 parameters characterizing these interactions.  
These parameters are treated as variables. In addition, there are two variables specifying the rotational temperature of the benzene molecule and the collision energy of Ar with benzene. 
The variable space is thus 10-dimensional. 
The collision lifetimes are computed using the classical dynamics method described in Ref. \cite{zhiying}. The GP model of the dependence of the collision lifetimes on these 10 parameters is then trained with results of 200 dynamical calculations as described in Section II above.  Figure \ref{figure:ar-benzene} compares the GP predictions with the dynamical calculations and confirms that a $10$-dimensional GP model can be trained with only 200 (order of $100 = 10 \times$ number of dimensions) calculations. The relative error of the GP predictions thus obtained is about 4 \%.

Figure \ref{figure:ar-benzene}  is an illustrative example showing the possibility of building a general GP model  giving a physical observable as a function of individual Hamiltonian parameters. One of the most important applications of such models is the evaluation of the uncertainties of the solutions of the nuclear Schr\"{o}dinger equation stemming from the errors of quantum chemistry calculations used to produce the underlying PES. This problem is discussed in the following subsection.

\subsection{Error bars of quantum dynamics results}

Assuming one can solve the Schr\"{o}dinger equation for nuclear dynamics numerically exactly, the results are still subject to uncertainties stemming from the errors of the $N$-dimensional PES $V(\bm r)$. There are two sources of errors. One is the inherent error of the quantum chemistry method used to produce the electronic energies. The other is the error of the analytical fit used to represent $V(\bm r)$ in a mathematical form suitable for solving the nuclear Schr\"{o}dinger equation. Although the errors of quantum chemistry calculations are not always easy to determine, they can be estimated by extrapolation to complete basis set results and by comparison with more sophisticated quantum chemistry calculations \cite{CBE}. The errors of the analytical fits can be determined by  validating the accuracy of the fit using the {\it ab initio} points not used for the production of the fit. 

Once the error of $V(\bm r)$ is determined, one needs to determine the effect of this error on quantum dynamics results. This has been attempted in multiple studies aiming to make predictions of scattering cross sections of molecules at low energies \cite{hutson1,hutson2,hutson3,jie,yura}. In these studies, the PES was multipled by a single scaling parameter and the effect of varying this parameter on the dynamics results was examined. However, scaling the entire PES by a single parameter does not change the anisotropy of the PES and 
does not account for the fact that different parts of the PES affect different dynamical results in a different way. For example, the elastic scattering cross sections are much less sensitive to the anisotropy of the PES than the probabilities of scattering that changes the angular momenta of interacting molecules.

The approach discussed in this subsection bears similarity to the Bayesian calibration of force fields in molecular dynamics simulations (see, for example, Refs. \cite{bayesian-calibration-1,bayesian-calibration-2,bayesian-calibration-3,bayesian-calibration-4,bayesian-calibration-5,bayesian-calibration-6,bayesian-calibration-7}). The goal of this work is either to quantify the error of the molecular dynamics simulations due to the uncertainty of the molecular force fields or to obtain the force field parameters best suited for the simulations of particular properties. 
One should note that the PESs used in quantum dynamics calculations are generally more complex than the molecular force fields used in classical dynamics. In addition,  quantum mechanics requires a more global representation of the molecular interactions and the quantum dynamics calculations are more time consuming than the classical dynamics calculations. This makes the problem discussed here more challenging in quantum dynamics. As discussed below and in Section IV.E, Gaussian processes allow one to address this challenge.

 In order to determine the effect of the uncertainty of $V(\bm r)$ on quantum dynamics results, it is necessary to vary different parts of the PES independently and examine the effect of these variations on the dynamical results. 
This is very challenging.  Consider, for example, a 6D PES given by an analytical fit with 10 parameters $\bm a = \{ a_1, a_2, ...a_{10} \}$ (in practice, the number of parameters is usually greater than 10). One can account for an arbitrary variation of the PES by varying each of $a_i$ individually. In order to account for the effect of this variation on the dynamics, one needs to solve the nuclear Schr\"{o}dinger equation each time one of these parameters is changed. If the calculations are performed on a simple grid of this 10-dimensional space with 10 points per $a_i$, this requires $10^{10}$ solutions of the nuclear Schr\"{o}dinger equation, clearly, an impossible task.

However, one can approach this as a Bayesian machine learning problem \cite{jie-prl}, described in the previous section. The variables of the ML model are the parameters $\bm a$. 
The model $y(\bm a)$ is provided by the Schr\"{o}dinger equation. 
The approach would entail solving the nuclear  Schr\"{o}dinger equation for $n$ random combinations of the parameters $\{ a_i \}$ to produce the training points $\bm y = [y_1, y_2, ..., y_{n} ]^\top$ and a GP model of $y(\bm a)$ trained by these solutions would provide the global dependence of the dynamical results on the individual coefficients $a_i$. If the dependence of $y$ on $\bm a$ is relatively smooth (as is expected to be the case), the number of training points required to produce an accurate GP model is known from ML literature to be on the order of $10 \times$(the number of dimensions in $\bm a$, 10 in our example) \cite{number-parameters}. Thus, an accurate dependence of the dynamical results on each of the coefficients $a_i$ can be obtained with only $\approx 100$ dynamical calculations, instead of $10^{10}$. This is illustrated in Figure \ref{figure:ar-benzene}.

\begin{figure}[!ht]
\centering
	\includegraphics[width=0.45\columnwidth]{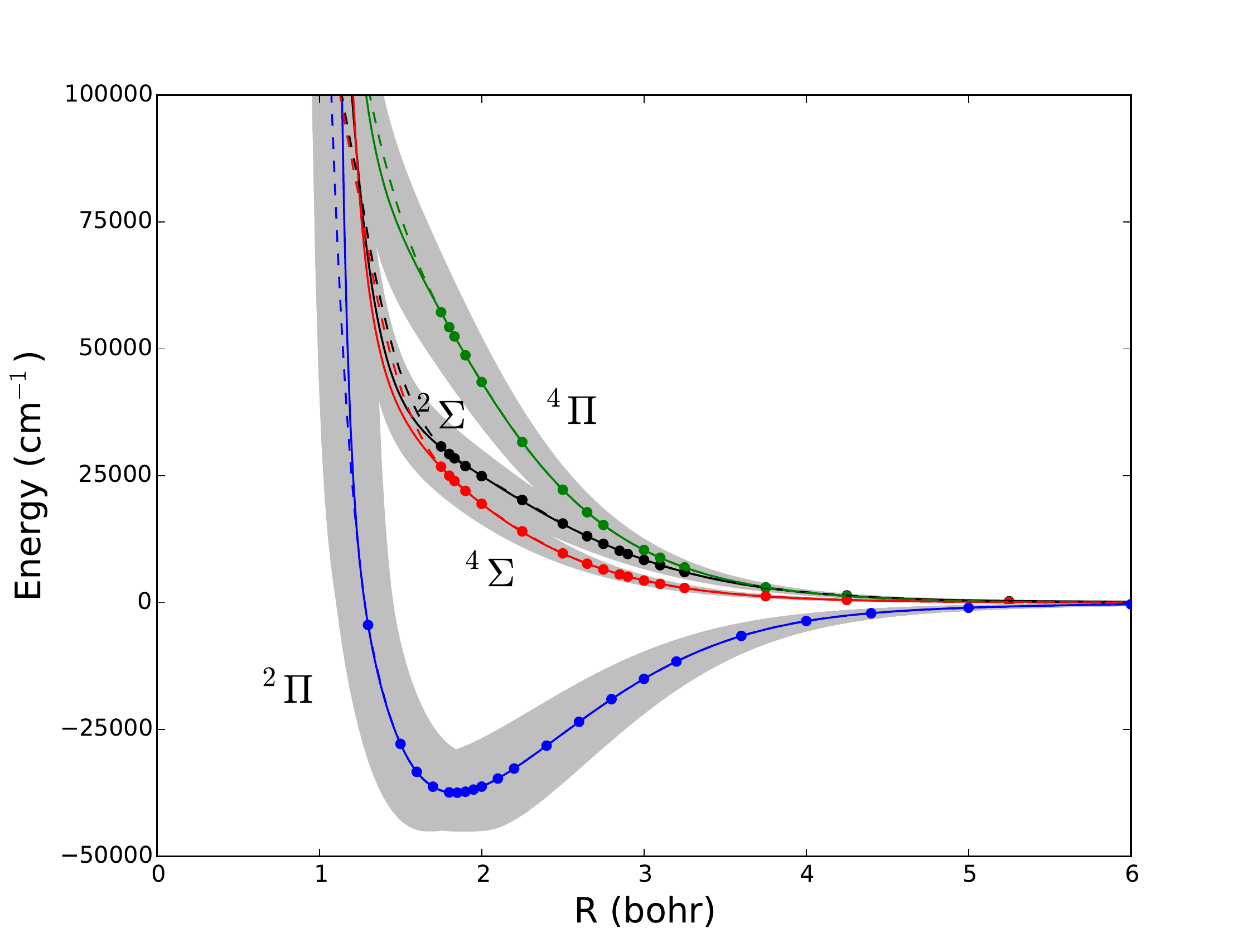}
	\includegraphics[width=0.45\columnwidth]{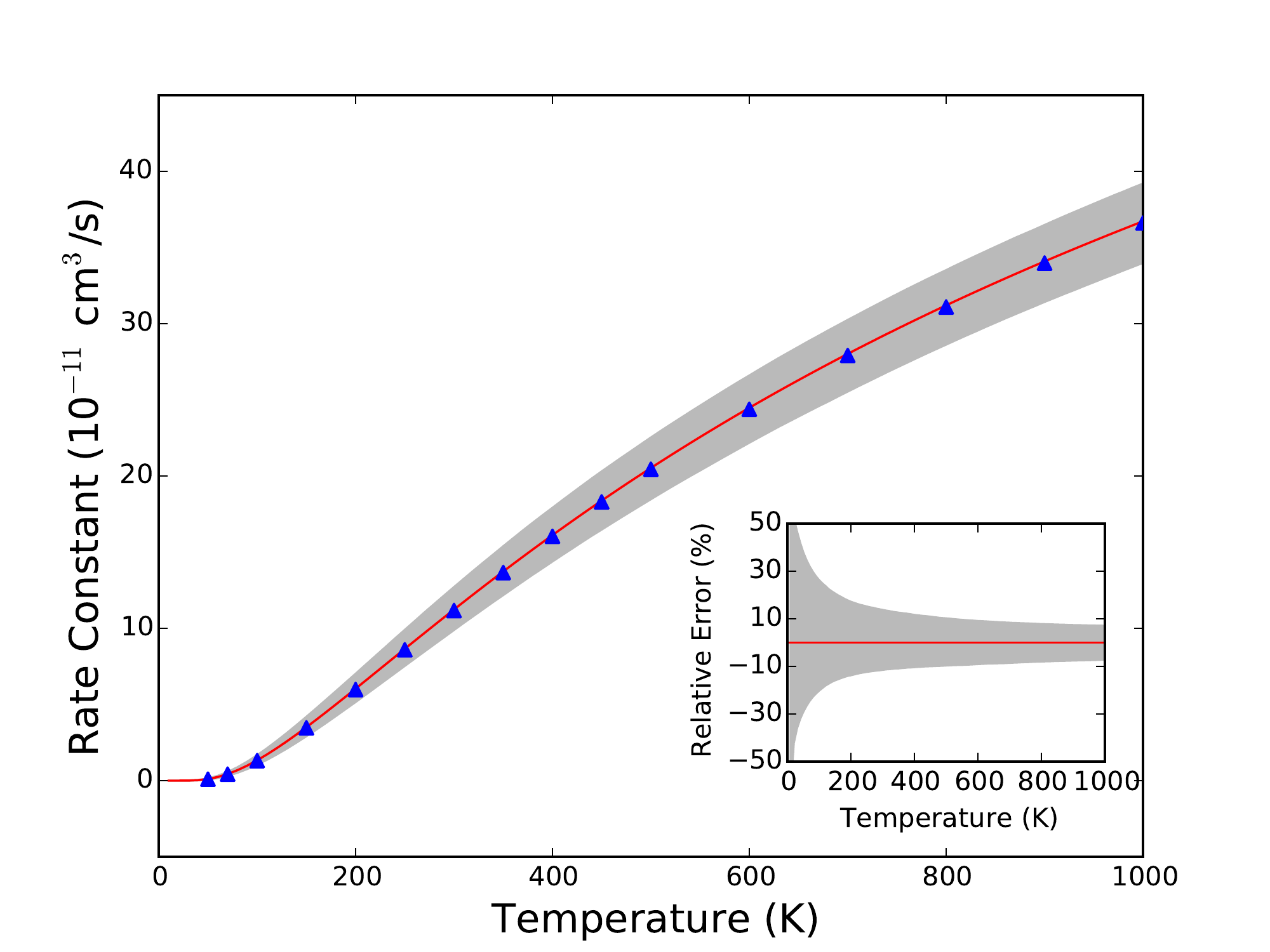}
	\caption{
	Left panel: The adiabatic interaction potentials determining the dynamics of fine structure transitions in collisions of O($^3P$)  with H atoms. The grey regions shows the assumed uncertainty of the potentials. This uncertainty propagates into the uncertainty of the collision rates. Right panel: The results of rigorous quantum scattering calculations  (triangles) with the grey area showing the uncertainty of the collision rate constants produced by the GP model. Figure adapted with permission from Ref. \cite{daniel-apj}. 
}
\label{figure:oh}
\end{figure}

Figure \ref{figure:ar-benzene} is an illustrative example that shows the possibility of constructing a general GP model $y(\bm a)$  giving a physical observable as a function of individual PES parameters. Once the GP of $y(\bm a)$ is trained, it can be used to examine the sensitivity of the dynamical results to each of the individual coefficients $a_i$ and calculate the uncertainty of the dynamical results associated with independent variation of each of $a_i$.  This procedure can be applied to any dynamical process, including non-adiabatic processes. Figure \ref{figure:oh} illustrates the application of this approach to determining the error bars of the rate constants for fine structure transitions in collisions of O($^3P$) with H. These rates are of significant importance for models of interstellar clouds. The fine structure transitions  in O($^3P$) - H collisions are determined by four adiabatic interaction potentials shown in the left panel of Figure \ref{figure:oh}. Given some ($R$-dependent) uncertainty to each of these potentials, a GP model was constructed in Ref. \cite{daniel-apj} based on solutions of the Schr\"{o}dinger equation as described above. This GP model was then used to obtain the uncertainty of the collision rate constants (shown in the right panel of Figure \ref{figure:oh}).

\subsection{Fitting PES with ML models}

Fitting multi-dimensional potential energy surfaces with analytical functions is a complex task required for any calculations of dynamical properties of molecules. 
 Although many different methods for fitting PES for polyatomic molecules have been developed over the past 50 years (see, for example, Refs. \cite{fitting-PES,general-fitting-1,general-fitting-2,general-fitting-4,general-fitting-5}), the development of efficient and universal fitting methods, which produce PES suitable for quantum dynamics calculations, is still a very active research field \cite{workshop}. This is primarily because, for polyatomic (multi-dimensional) systems, PES often exhibit nontrivial dependence on the configuration space coordinates that cannot be captured by standard sets of analytical functions and that are often system specific. 

Since ML aims to build efficient models of unknown functions, it is well suited for constructing PES for polyatomic molecules \cite{ML-for-PES}. 
Fitting PES with ML models has indeed been gaining popularity in recent years. There are two trends in the literature on fitting PES with ML models. One is towards developing NN fits tailored for specific quantum dynamics calculations. Other studies aim to exploit kernel-based methods, including GP models,  for the automated construction of complex PES. 
It is, therefore, useful to consider separately and compare the NN models and the GP models of PES. Table I summarizes the relative advantages and drawbacks of the two types of ML methods for fitting PES.

\begin{table}
\caption{Relative advantages and disadvantages of the GP and NN models for fitting PES for polyatomic molecules. The number of potential energy points used to construct the model is denoted by $n$. 
\label{table:GPvsNN}}
\begin{tabular}{llllll}
\hline\hline
 \hspace{5.cm} &  GP models of PES & NN models of PES \\
\hline
Training complexity           &  Scales as ${\cal O}(n^3)$ & Variable, can be fast \\
Speed of evaluation &  Scales as ${\cal O}(n)$ & Very fast, independent of $n$ \\ 
Accuracy                  &  Low error (easy)  & Low error (easy) \\
Extreme accuracy    & Extremely low error (difficult) & Extremely low error (easy) \\
Required $n$ & Small & Large \\
Construction of fit & Automated (with kernel chosen) & Requires manual work \\
Extension to more dimensions & Automated  & Requires manual work \\ 
Extension to other molecular systems & Automated  & Requires manual work \\
Overfitting & Rare, only if complex kernels & Likely \\
Final model & Non-parametric, numerical & Analytical, can be tailored  \\
& vector - vector product & for specific dynamics software \\
\hline
\hline
\end{tabular}
\end{table}

 The use of NNs for fitting PES was introduced by the work of Manzhos and Carrington \cite{NNs-for-PES, NNs-for-PESa} and Behler and Parrinello \cite{NNs-for-PES-1a, NNs-for-PES-1b, NNs-for-PES-1c},
  and is currently exploited by many authors \cite{NNs-for-PES-4, NNs-for-PES-2, NNs-for-PES-3, NNs-for-PES-5,NNs-for-PES-6}, often because NNs allow one to construct the sum-of-product representation of PESs suitable for quantum dynamics methods, such as ones based on the multi-configuration time-dependent Hartree method \cite{MTDH}. NNs definitely offer many advantages for fitting PES compared to other fitting methods, in general, and GPs, in particular.  
  One is the speed of the evaluation of the resulting PES, once the NN is trained. Another advantage is the possibility of constructing PES with extremely low fitting error due to the flexibility of the fit. NNs can be used to construct fits with permutation symmetry built in \cite{hua-guo-permutation-NN}. However, NNs require substantial user experience informing the choice of the NN architecture and the initial choice of the NN parameters. Training a NN fit of a PES, especially, for complex, multi-dimensional systems thus often requires manual work.  One should also be careful to avoid overfitting, a common problem in applications of NNs. 

The idea to use kernels to represent PES was introduced by the work Rabitz and coworkers \cite{general-fitting-2, rabitz-1,rabitz-2,rabitz-3}. 
GPs have been introduced to the molecular dynamics community more recently, first as a method for fitting the force fields for classical dynamics simulations  \cite{ML-for-MD-1, ML-for-MD-3, gp-1,gp-2,gp-3}, and then as a method
for fitting global PES that could be used in quantum dynamics calculations of scattering or spectroscopic properties
\cite{jie-jpb, gp-for-PES-2, gp-for-PES-3, gp-for-PES-4, gp-for-PES-5, gp-for-PES-6, gp-for-PES-7, gp-for-PES-8, gp-for-PES-9}. The GP models are significantly slower to evaluate than the NN fits because the prediction of the potential energy is in the form of the vector - vector product (\ref{GP-mean}) that needs to be evaluated numerically. The number of training points (i.e. the size of the vectors in Eq. \ref{GP-mean}) increases with the number of degrees of freedom. On the other hand, GPs require much fewer training points than NNs \cite{rodrigo-bo}. 
 In Ref. \cite{rodrigo-bo}, it was found that an accurate GP model of the global PES for the quantum scattering calculation of the probabilities of the reaction of OH with H$_2$ can be constructed with less than $300$ potential energy points. The same surface was previously constructed with a NN fit using $\approx 17,000$ potential energy points \cite{NNs-for-PES-5}. 
The relative performance of GPs vs NNs for the construction of global PES for polyatomic molecules was examined in Ref. \cite{gp-for-PES-4}. 

GPs offer two fundamental advantages for fitting PES, which could be exploited for new applications. First and foremost, the interpolation of a PES by GPs 
does not require any manual work. The same code can be applied to obtain the global surface for any molecular system with any number of degrees of freedom. The accuracy of the surface can be improved by simply adding more potential energy points to the training set.   
As illustrated in Ref. \cite{jie-jpb}, the accuracy of the GP model of PES monotonously decreases with the number of training points and there is usually no additional work required to avoid overfitting when building PES with GPs. This is likely because global PES for polyatomic molecules can be constructed with simple kernels  \cite{jie-jpb, gp-for-PES-2, gp-for-PES-3, gp-for-PES-4, gp-for-PES-5, gp-for-PES-6, gp-for-PES-7, gp-for-PES-8, gp-for-PES-9}.  
Different surfaces can also be constructed automatically by the same program, using different sets of potential energy points.
Second, GP models - as a consequence of the Bayesian approach -- offer not only the fit of the PES but also the uncertainty of the fit in the form of Eq. (\ref{GP-variance}). As described in Section II above, this uncertainty can be exploited for Bayesian optimization. The following two subsections discuss examples of new applications made possible by these properties of GP models. 


\subsection{Learning from Machine Learning}

When faced with the problem of calculating a PES for a new molecular system, one must choose how to place the potential energy calculations in the configuration space. 
We refer to this as the sampling problem.  Although it is feasible to perform thousands and even tens of thousands of accurate quantum chemistry calculations for most molecules, 
sampling is a significant problem for high-dimensional systems with unknown PES landscape. This is especially true if the PES includes unusual geometrical features, such as conical intersections. There have been several recent attempts to address this problem with ML \cite{gp-for-PES-2, gp-for-PES-6, rodrigo-bo}. The possibility of automated construction of PES with GPs
combined with Bayesian optimization (BO) suggest a ML method of obtaining the ideal sampling scheme. 

This problem can be best formulated with reinforcement learning (RL). The target of the RL algorithm is the most accurate PES obtained with as few potential energy points as possible. 
The result is the most efficient sampling scheme. Within this formulation, one asks the question: `Given a PES constructed with $n$ potential energy points, where in the configuration space should the next {\it ab initio} point be placed to maximize the accuracy of the PES?' GP regression offers a very unique way to answer this question because it is known (as illustrated in Figure \ref{figure:BML}) that any GP model with $n \rightarrow \infty$ training points produces the most accurate surface. This is a consequence of the Bayesian approach. If one uses GP regression for representing a PES, the question above can be reformulated thus: `How does one get to the $n \rightarrow \infty$ limit with as few steps as possible?' As illustrated in Ref. \cite{rodrigo-bo}, this question can be answered by RL with the help of BO. 

The specific algorithm can be formulated as follows. The procedure begins with a small number  $n$ (say, $n=100$ for a 6D space) {\it ab initio} points placed randomly in the configuration space. These points are used to train a GP model of the PES denoted by ${\cal G}(n)$. Given ${\cal G}(n)$, one adds another potential energy point in the configuration space and trains another GP model ${\cal F}(n+1)$ based on $n+1$ points. This point is added at random locations of the configuration space, thus ${\cal F}(n+1)$ becomes a function of the coordinates of the added point $\bm x$. BO -- as described in Section II -- can then be used to {\it maximize} the difference $|{\cal F}(n+1) - {\cal G}(n)|$ to produce the most optimal location $\bm x^{\rm opt}$ for the added potential energy point. With this procedure iterated, BO yields the most optimal distribution of energy points in the configuration space. 
This was illustrated in Ref.  \cite{rodrigo-bo}, where BO was used to build PES producing accurate quantum dynamics results with as few potential energy points as possible.  
Although this has not yet been done, one can repeat these calculations for a variety of molecular systems and derive the most optimal (potentially universal) sampling scheme. 
Thus, one could learn the optimal sampling scheme from machine learning. 

\subsection{Inverse scattering problem}

Consider a chemical reaction of two molecules. In order to predict the reaction probabilities accurately, one needs to perform quantum dynamics calculations on a global PES. As mentioned above, the traditional approach begins with building a global PES independently of the quantum dynamics calculations. However, not all parts of the global PES are equally important for the outcome of specific quantum dynamics calculations. In terms of the potential energy calculations and the amount of information offered by the PES, the traditional approach is, therefore, almost always an overkill. Moreover, different reaction processes can be more or less sensitive to different parts of the PES.  A global PES equally accurate in the entire configuration space does not provide information about which part of the configuration space is sampled by the quantum reaction process under study.  

Ref.  \cite{rodrigo-bo} shows that the unique properties of GPs allow one to invert the problem and build the global PES by placing the potential energy points only in 
the part of the configuration space that is important for a specific quantum reaction process. This is achieved by the following iterative process: 

\begin{minipage}{0.4\columnwidth}
\centering
 \begin{tikzpicture}[->,scale=.7]
   \node (i) at (90:1cm)  {\small A};
   \node (j) at (-30:1cm) {\small F};
   \node (k) at (210:1cm) {\small D};
   \draw (70:1cm)  arc (70:-10:1cm);
   \draw (-50:1cm) arc (-50:-130:1cm);
   \draw (190:1cm) arc (190:110:1cm);
   
\end{tikzpicture}
\end{minipage} \hspace{4.cm} (29)
\setcounter{equation}{29}

\noindent
where step A produces  a potential energy point by an {\it ab initio} quantum chemistry calculation {\it or} moves a particular energy point along the energy axis, step $F$ produces a global model of the PES by GP regression (\ref{GP-mean}) and step D produces the reaction probabilities by solving the nuclear Schr\"{o}dinger equation with this PES. Moving the energy points along the energy axis allows one to account for the error of quantum chemistry calculation and ensures that step D can produce results in agreement with any desired outcome \cite{rodrigo-bo}. Each time an energy point is added or moved in step A, the global surface produced in step F becomes different and the outcome of the dynamical calculation in step D changes. 

Given some experimental measurements of quantum dynamics, one can then apply BO to the feedback loop (29) in order to minimize the difference between the outcome of step D and the given physical data. In principle, one could implement the feedback loop (29) without ML. However, in practice it would be impossible because every iteration of this feedback loop requires 
the generation of a new, global PES in step F and solving the nuclear Schr\"{o}dinger equation in step D.  The feedback loop relies on the ability of GP regression to produce a new global surface at each iteration automatically, without any manual work. BO ensures that the iterative process (29) converges very fast and requires very few scattering calculations. 

\begin{figure}[!ht]
\centering
	\includegraphics[width=0.45\columnwidth]{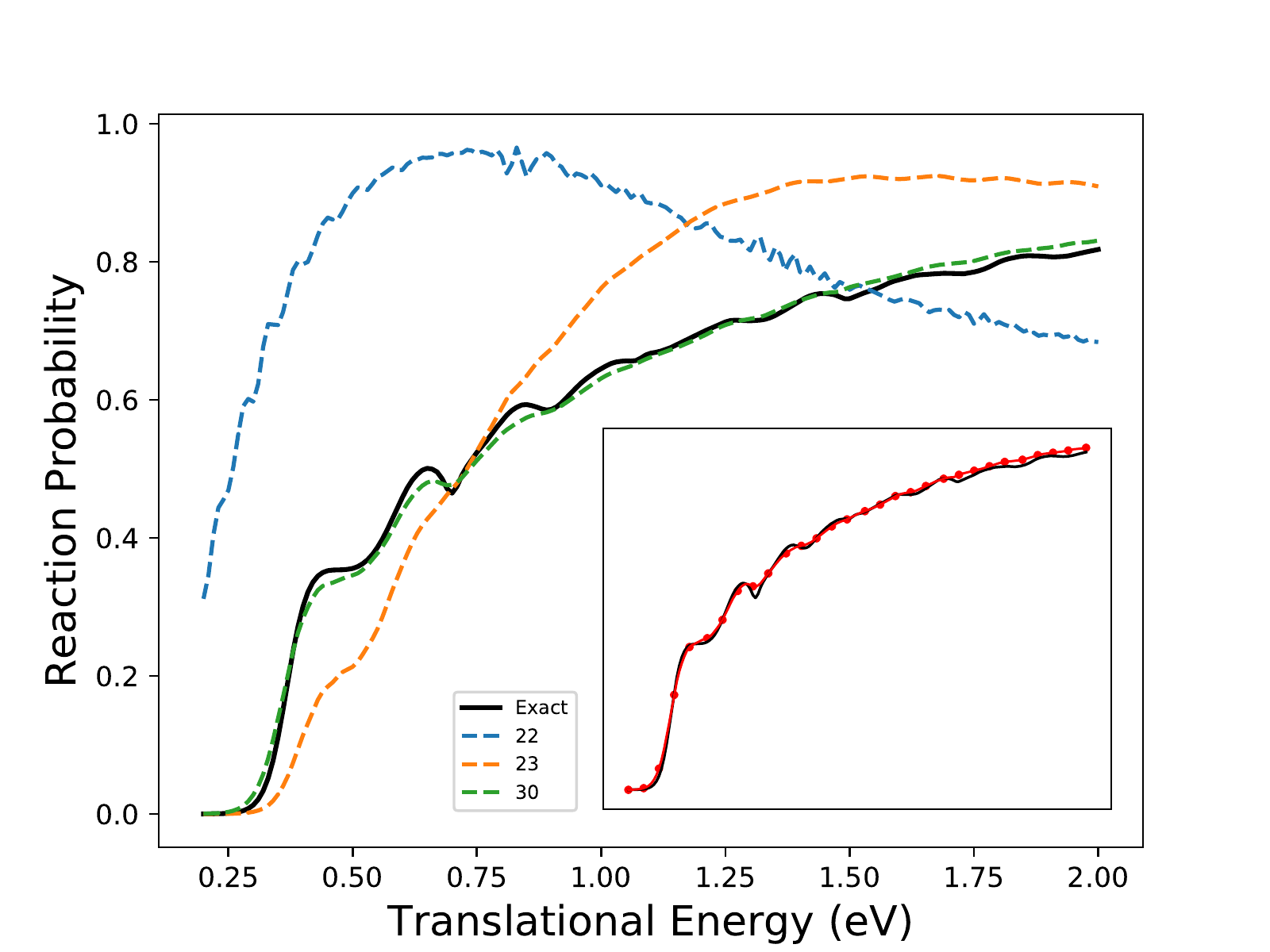}
	\includegraphics[width=0.45\columnwidth]{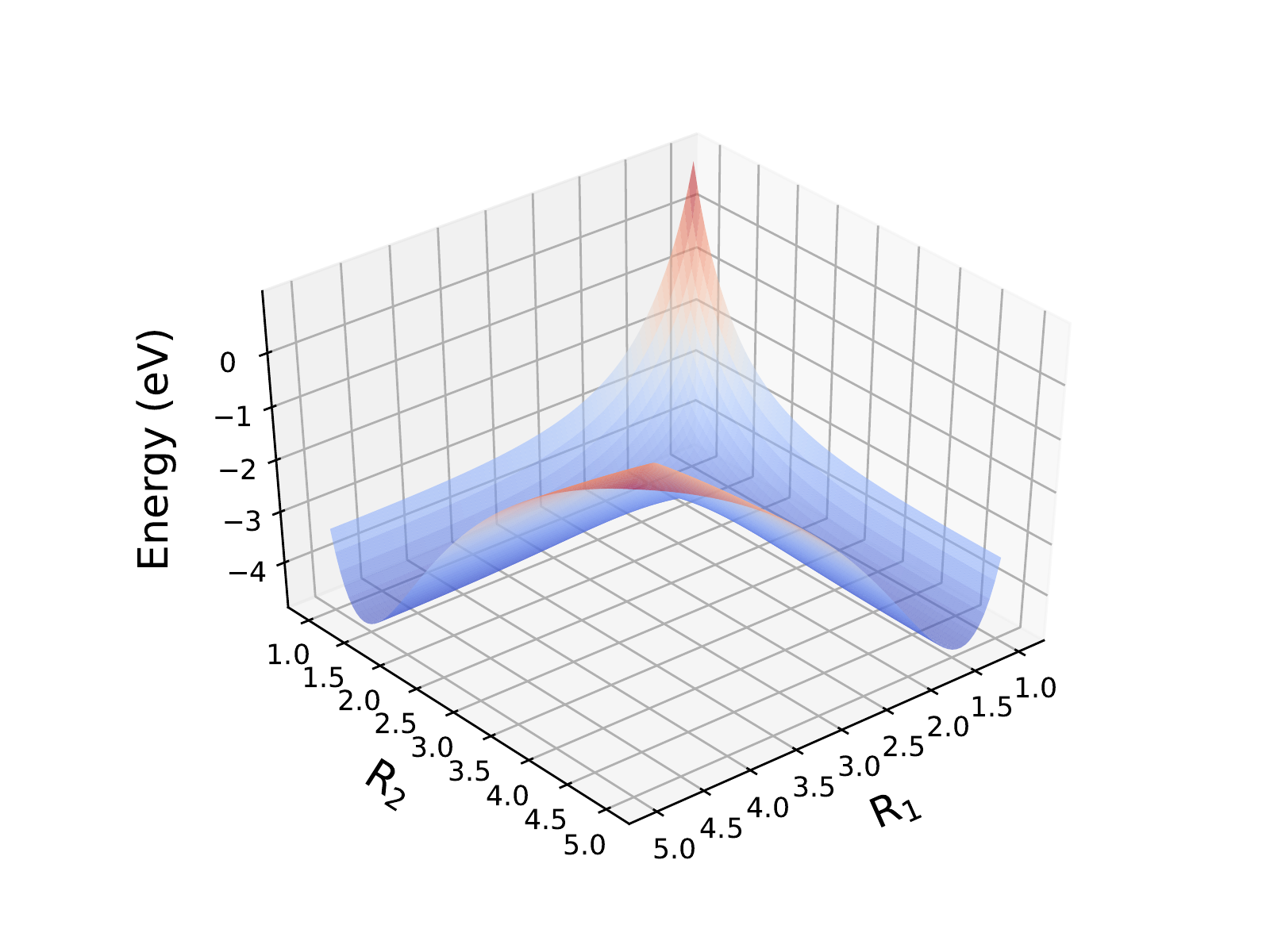}
	\caption{
	Left panel: The total probability for the H$_2$ + H $\to$ H + H$_2$ reaction as a function of the collision energy. The black solid curve -- calculations from \cite{h3} based on the surface with 8701 ab initio points from \cite{h3-pes}. The dashed curves -- calculations based on the GP PES obtained with 22 ab initio points (blue); 23 points (orange), 30 points (green) and 37 points (inset). The points 23--37 are drawn by the BO algorithm from the PES in \cite{h3-pes}. The RMSE and maximum error of the results with 37 points are 0.009 and 0.028, respectively. Right  panel: the GP model of the PES for the H$_3$ reaction system constructed with 30 potential energy points. $R_1$ and $R_2$ are the distances between atoms 1 and 2 and atoms 2 and 3, respectively. Figure adapted with permission from Ref. \cite{rodrigo-bo}. 
}
\label{figure:h3-BO}
\end{figure}

These points are illustrated in Figure \ref{figure:h3-BO}. To generate the results in this figure, we consider  the chemical reaction  H + H$_2$ $\rightarrow$ H$_2$ + H.  A global PES for this reaction was constructed by an analytical fit to 8701 potential energy points in Ref. \cite{h3-pes}. The quantum dynamics calculations of reaction probabilities 
based on this full PES were performed in Ref. \cite{h3}. The results of Ref. \cite{h3} are shown in Figure  \ref{figure:h3-BO} by the black solid curve. The first question we address is: ``Is it possible to obtain the results shown by the black solid line using a small subset of the 8701 potential energy points?'' To answer this question, we start with $n = 22$ potential energy points randomly distributed in the 3D configuration space and implement the feedback loop (29) to add one point from the set of 8701 points at each iteration. As shown in Figure \ref{figure:h3-BO}, BO converges the results of step D in the feedback loop to the black solid curves with only 8 iterations (which require 360 quantum scattering calculations). The inset shows the energy dependence of the reaction probabilities (red symbols) obtained with a PES represented by a GP model with only 37 {\it ab initio} points (instead of 8701!). The surface based on the 37 potential energy points is shown in the right panel of Figure \ref{figure:h3-BO}.


Ref. \cite{rodrigo-bo} shows that the convergence of the BO is similarly fast when applied to a more complex reaction system. For example, in the case of the reaction OH + H$_2$ $\rightarrow$ H$_2$O + H, the feedback loop starting with  280 potential energy points randomly distributed in the six-dimensional configuration space produces accurate reaction probabilities after 10 iterations, leading to accurate dynamical results based on a GP model of the global PES trained by only 290 points, which should be compared to $\approx 17,000$ {\it ab initio} points used for the NN fit of the same PES in the previous work \cite{NNs-for-PES-5}.

This subsection illustrates that the combination of GP modelling of global reactive PES with Bayesian optimization allows one to construct accurate PES with a very small number of potential energy points. By construction, these PES, when inserted into the Schr\"{o}dinger equation, lead to accurate dynamical results for a particular reaction process. 
The PES thus obtained contains a very useful information about which part of the configuration space is the most important for the reaction process under study and thus offers information that can be used to elucidate the mechanisms of the reaction processes. The feedback loop (29) thus again offers the possibility of learning microscopic physics of molecular reactions from machine learning.

\subsection{Accurate models based on approximate results}

It is difficult to solve Eq. (\ref{the-differential-equation}) exactly for complex molecular systems in a wide range of Hamiltonian parameters. The computation effort in quantum dynamics is often lowered by applying either decoupling approximations, which reduce the number of coupled differential equations in  (\ref{the-differential-equation}), or by freezing some degrees of freedom.
The computation effort may also be reduced by replacing the quantum treatment of some (or all) degrees of freedom with the classical treatment. 
These approximations introduce additional errors into predictions of dynamical results. ML in general \cite{ML-accuracy, ML-accuracy-2} and Bayesian ML, in particular, offers methods to construct models that `learn' these errors and correct the approximate results.

\begin{figure}[!ht]
\centering
	\includegraphics[width=0.55\columnwidth]{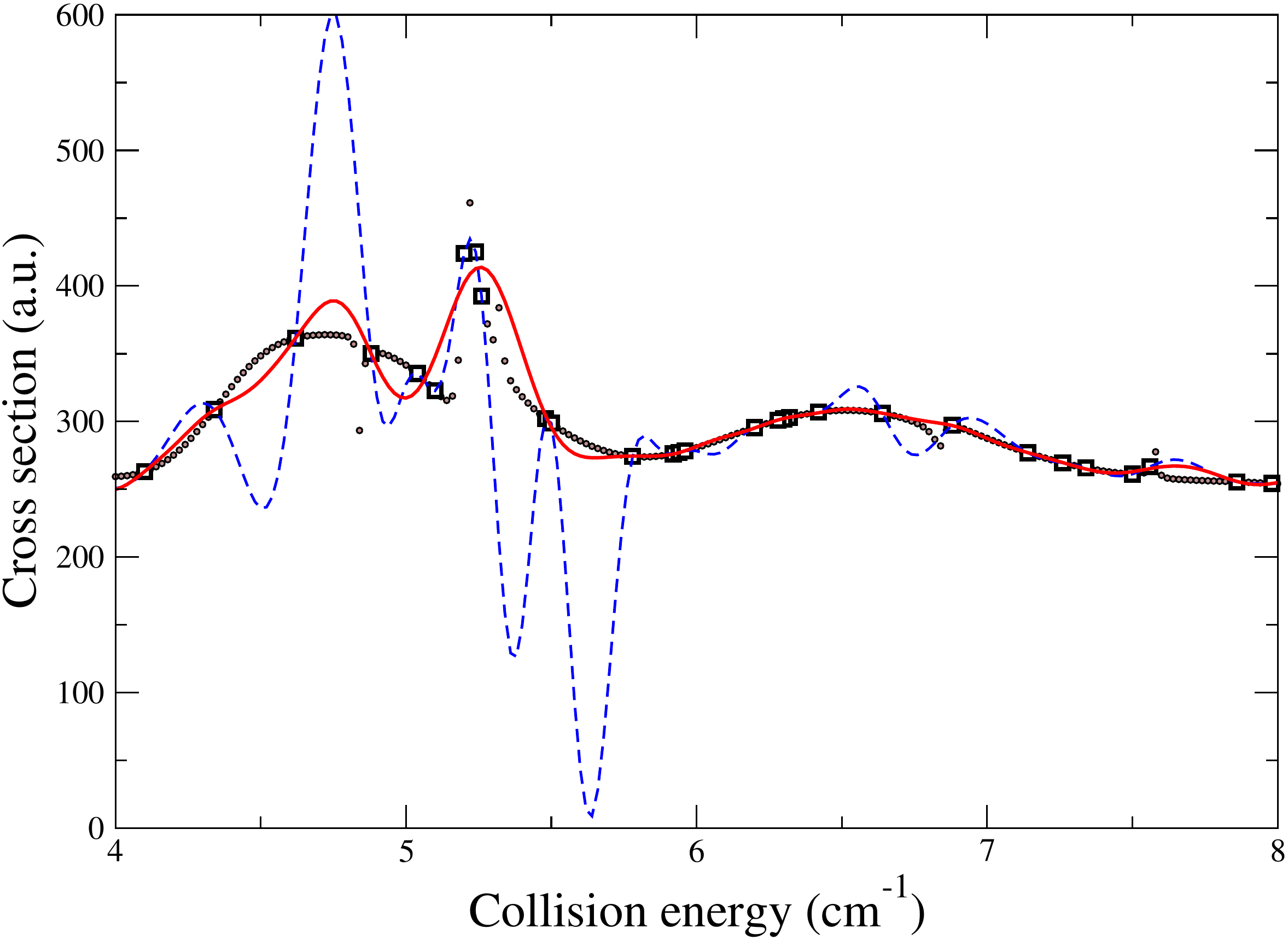}
	\caption{The model (\ref{double-GP-eq}) of the cross sections for C$_6$H$_6$ - He collisions based on a small number of quantum scattering calculations represented by squares and a large number of approximate classical trajectory calculations (not shown). The model results are shown by the red curve and should be compared with the fully quantum results (circles) that are not used for training the model. The model trained by the quantum dynamics results without any classical trajectory calculations is shown by the blue curve. 
 Figure adapted with permission from Ref. \cite{jie-prl}. 
}
\label{figure:double-GP}
\end{figure}

Consider some observable computed at $n$ values of the Hamiltonian parameters
by the rigorous quantum dynamics method and at $m$ values of the Hamiltonian parameters by some approximate quantum dynamics methods.  
The rigorous results are $\bm y = \left [y_1, y_2, ..., y_n \right]$ and the approximate results are  $\bm z = \left [z_1, z_2, ..., z_m \right]$. 
Because the approximate calculations are easier, $m \gg n$. Given this information, one can build a model of the accurate results $\bm y$ in a wide range of the Hamiltonian parameters as follows \cite{double-model}: 
\begin{eqnarray}
{\cal E(\cdot)} = \rho {\cal F(\cdot)} + {\cal G}(\cdot) + \varepsilon, 
\label{double-GP-eq}
\end{eqnarray}
where ${\cal F(\cdot)}$ is a ML model trained by the approximate results $\bm z$, and ${\cal G}(\cdot)$ is trained to model the difference between the approximate $\bm z$ and rigorous $\bm y$ results at those few $n$ points given by the rigorous calculations. The advantage gained by training the model with a large number of approximate results is illustrated in Figure (\ref{figure:double-GP}).

This procedure is rather general and can be applied to any theoretical chemistry result obtained with a large number of approximate calculations and a small number of rigorous results, as long as the approximate and  exact results are correlated. In Ref. \cite{gp-for-PES-9}, a similar approach was used to produce accurate potential energy curves for the N$_2$ molecule using a combination of quantum chemistry results at different levels of theory. Used as in Ref. \cite{gp-for-PES-9}, this method can be used to evaluate the basis set errors as well as the systematic error of the quantum chemistry method.

The exact values $\bm y$ can be the results of rigorous quantum calculations as above or the results of experimental measurements. 
If both $\bm y$ and $\bm z$ come from theoretical calculations of different accuracy, $\varepsilon$ should generally be set to zero. 
If $\bm y$ come for an experiment, ${\cal G}(\cdot)$ models the systematic errors of the theoretical method and $\varepsilon$ accounts for the noise in the experimental data. 
The technique described in this section should be broadly viewed as the interpolation of a small number of accurate results assisted by a large number of approximate results. 
It would also be desirable to develop techniques that could use the accurate results in a small range of the Hamiltonian parameters and make predictions outside the range of the training data by extrapolation. While ML has not yet been applied for this purpose in molecular physics, there are some encouraging results from applications in condensed-matter physics indicating the  feasibility of physical extrapolation with ML. This problem is discussed in the following subsection.

\subsection{Extrapolation of quantum observables}

\begin{figure}[!ht]
\centering
	\includegraphics[width=0.55\columnwidth]{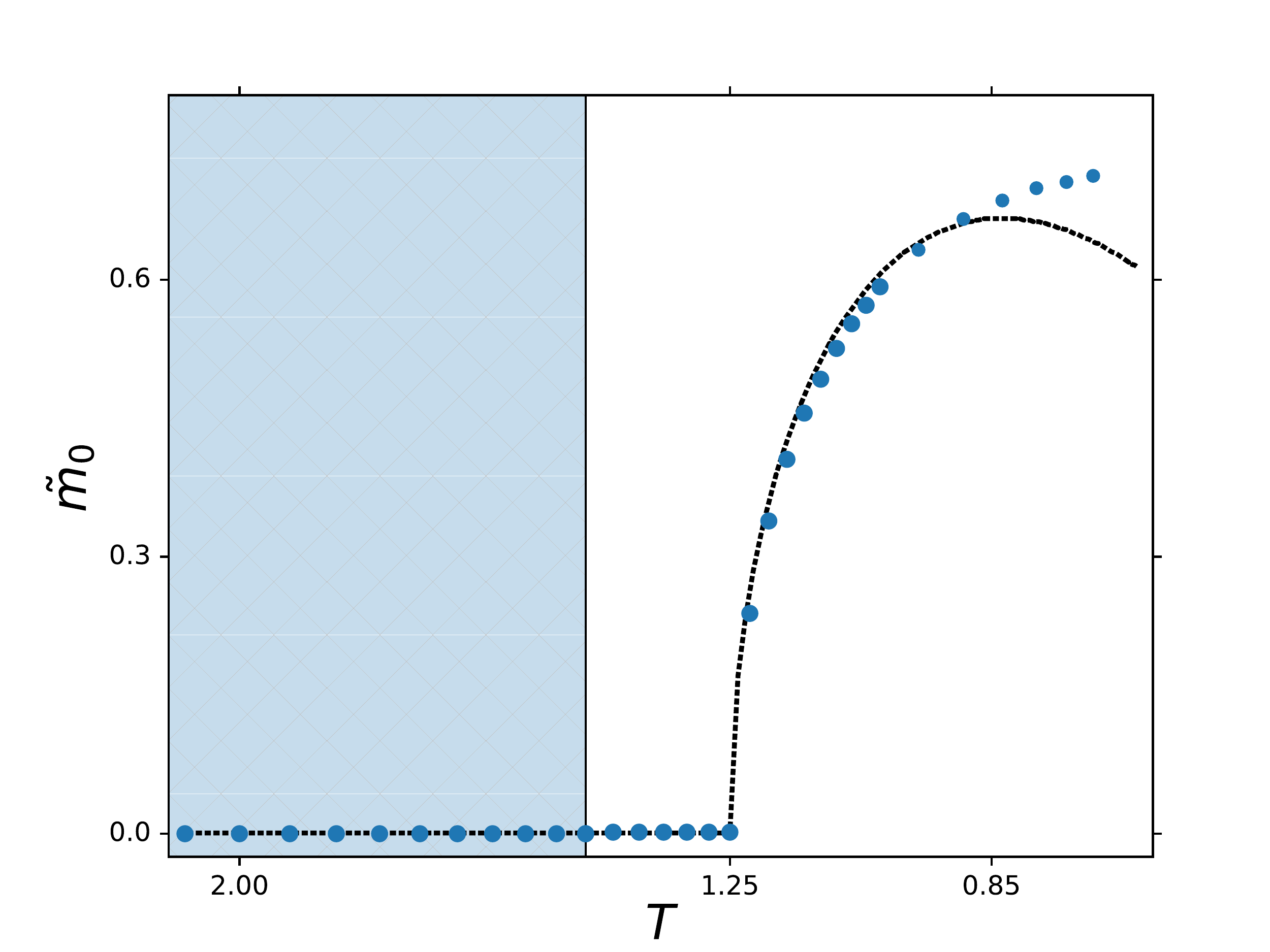}
	\caption{Dependence of the order parameter $\tilde m_0$ for the Heisenberg spin model on temperature $T$ for an infinite one-dimensional spin chain. The symbols are the predictions of the GP model. The dashed curve shows the results computed from the analytic formula for the free energy obtained in the mean field approximation. 
	Note that the dashed curve is not used for training the GP model. The GP model is trained by the two-dimensional dependence of the free energy on temperature and magnetization at temperatures in the shaded region. This two-dimensional dependence of the free energy is then extrapolated to temperatures outside the shaded region and the order parameter is 
calculated from the extrapolated surface as the minimum value of the magnetization at a given temperature. The evolution of the free energy with temperature is smooth. 
 Figure adapted with permission from Ref. \cite{extrapolation-0}. 
}
\label{figure:extrapolation}
\end{figure}

Consider a quantum system described by a Hamiltonian that is a function of multiple parameters.  In the context of scattering theory, these parameters are, for example, the total energy or the total angular momentum of the system.  Solving the Schr\"{o}dinger equation for a range of these parameters produces the values of a desired observable $\bm y = \left [  y_1, y_2, ..., y_n \right]^\top$, which can be used to train ML models as discussed before. These ML models, whether coming in the form of NNs or GPs, are well-controlled within the range of the Hamiltonian parameters used for training. For example, the accuracy of GP predictions within the range of the training data is guaranteed to decrease as $n$ increases. 

  In the present section, I consider the possibility of using ML models to make predictions at the values of the Hamiltonian parameters outside the range used for training the models. 
As discussed in Section II.C, computer science literature \cite{extrapolation-1, extrapolation-2} and an application of this method to predictions of quantum phase transitions \cite{extrapolation-0} indicate that the predictive power of GPs can be enhanced by increasing kernel complexity to maximize the Bayesian information criterion (\ref{BIC-eq}). 

Figure \ref{figure:extrapolation} presents an example of the predictive power of a GP model with a kernel selected by the BIC \cite{extrapolation-0}. 
In this example, a GP model is trained by the dependence of the free energy of the one-dimensional Heisenberg spin model on two variables: temperature and magnetization. The value of magnetization at the minimum of the free-energy curve is the order parameter, which characterizes the paramagnetic and ferromagnetic phases of the system. The GP model is trained as described in Section II.C at temperatures in the shaded region of Figure \ref{figure:extrapolation}. The free energy is then extrapolated to temperatures outside the shaded region and the order parameter is computed from the extrapolated curves. The order parameter thus predicted is shown in Figure \ref{figure:extrapolation}  by symbols. The results based on the analytic mean-field calculation are shown by the dashed curve.

It is striking to see from Figure \ref{figure:extrapolation} that the GP model captures the physical evolution of the order parameter with temperature. This illustrates that the Bayesian information criterion (\ref{BIC-eq}), as suggested in computer science  \cite{extrapolation-1, extrapolation-2, bic}, can indeed be used to guide the selection of a physical model. This is important particularly in condensed-matter physics, where many quantum models can be numerically solved or accessed by experiments only in a limited range of Hamiltonian parameters. The extrapolation with GPs may in such cases be the only method to explore the properties of quantum systems at some values of the parameters. 
This method also holds great potential for applications in quantum molecular dynamics.  For example, the extrapolation technique could be combined with the numerical algorithms for 
solving either the time-dependent or time-independent Schr\"{o}dinger equation potentially allowing one to find accurate numerical solutions with much larger integration steps than in conventional methods. Another application could be the basis set extrapolation, commonly exploited in quantum chemistry, but less in quantum dynamics.

However, much work remains to be done to make extrapolation with GPs a useful tool for molecular physics. In particular, the generality of the prediction method, the reliability, the typical and maximum accuracy of the extrapolation predictions remain unknown. Can accurate PES be built by extrapolation with ML? Can the solutions of the Schr\"{o}dinger equation for molecular systems be extrapolated with accuracy better than provided by efficient approximate dynamical methods? Can extrapolation with GPs provide enough accuracy to accelerate the numerical algorithms for solving coupled differential equations? These important questions remain open.

\section{Conclusion}

This article has attempted to argue that combining quantum dynamics calculations of molecular properties with machine learning opens up new research opportunities. 
By providing examples, I have shown that ML offers the possibility to address new research questions that would be difficult or impossible to consider without ML.  
In order to take advantage of ML, it is useful to approach quantum molecular dynamics in somewhat unconventional ways.  
For example, instead of viewing atoms as undergoing dynamics on a given PES, Bayesian ML allows one to formulate the problem as the Schr\"{o}dinger equation with a non-parametric distribution of potential energy surfaces that becomes conditioned by the desired dynamical properties. This makes it possible to address questions such as ``What is the best PES for a particular outcome of a microscopic interaction of molecules?'' or ``What part of the configuration space is sampled by a particular quantum dynamics process?''. This formulation can be extended to a wide range of problems and can be applied, for example, to determining the microscopic details of molecule - surface interactions from the results of surface spin echo experiments \cite{gil}. 

One particular formulation of quantum dynamics advocated here is in the form of a ML simulator of the Schr\"{o}dinger equation. 
If combined with the Bayesian statistics, such a simulator allows one to obtain not only the quantum predictions but also the error bars of the dynamical results associated with an uncertainty of inputs (such as the PES) into the Schr\"{o}dinger equation. Predicting the uncertainties of the Schr\"{o}dinger equation solutions is as important as obtaining the solutions themselves. Such simulators can also be used to explore the sensitivity of the Schr\"{o}dinger equation solutions to any parameters in the Hamiltonian, thus offering mechanistic insights into molecular dynamics. For example, one can analyze the sensitivity of the dynamical results to a particular part of the underlying interaction potential. 

One can also determine the sensitivity of the observables to the collision energy as well as the basis sets used in the calculations. The sensitivity of the dynamical results to the collision energy or to the basis sets can be used to provide information about the range of the Hamiltonian parameters, where the Schr\"{o}dinger equation must be solved on a denser grid or with larger basis sets. For example, the presence of scattering resonances generally makes the dynamical properties more sensitive to the Hamiltonian parameters. The sensitivity analysis based on the ML simulators described here could be used to identify the parameter ranges more or less affected by resonances. At the same time, if the ML simulators are insensitive to the basis set size in a particular range of collision energies, external field strength or total angular momenta, one can perform rigorous dynamical calculations in this range of the Hamiltonian parameters with a smaller basis.

Machine learning, and in particular the Bayesian approach based on GPs, can automate many of the calculations of significant importance to quantum molecular dynamics. The most important example is the automated construction of PES with GPs. It has been shown that GPs can be used to build models of complex, multi-dimensional PES with non-trivial geometrical features. 
Other than the choice of the mathematical form of the kernel functions (which is not unique and can also be done automatically), the process of training a GP model of a PES does not require any manual work. This is particularly important for applications, where multi-dimensional surfaces must be constructed multiple times. Moreover, ML models of PES are flexible and do not rely on any particular functional dependence of the surface. As such, they can be applied to rather complex systems with unknown geometrical features. 

As illustrated, GP models allow one to design algorithms based on Bayesian optimization.  In quantum dynamics, Bayesian optimization can be used to provide reference data. For example, one can use Bayesian optimization to determine the most efficient distribution of potential energy points in the configuration space, yielding the global PES with the least number of energy points. The sampling schemes for other molecules can be modelled after such sampling distributions. As illustrated, Bayesian optimization can also be used to solve the inverse scattering problem.

ML models can also be trained to analyze the difference between rigorous calculations and approximate dynamical results. This is especially useful when rigorous calculations are extremely expensive and approximate results have unknown errors. In such cases, the number of rigorous results is usually not sufficient to train an accurate ML model directly and an accurate interpolation of rigorous results can be obtained by combining a ML model of the approximate calculations and a ML model trained by the difference of the rigorous and approximate results. One can envision numerous applications of this approach, ranging from constructing accurate multi-dimensional PES based on low-level quantum chemistry calculations, to obtaining accurate dynamical results in a wide range of the Hamiltonian parameters with approximate dynamical methods. For example, one can reduce the number of coupled differential equations in Eq. (\ref{the-differential-equation}) by ignoring the Coriolis couplings in the Hamiltonian  \cite{cs-1,cs-2,cs-3,cs-4}. However, the accuracy of this approximation may vary depending on the Hamiltonian parameters as well as the presence or absence of peculiar scattering features, such as resonances.  The approach based on the combined ML model of a small number of rigorous calculations and a large number of approximate dynamical results may in this case be used to produce the dynamical observables at the same level of accuracy as the full dynamical calculations.

Finally, I have presented evidence that GP models may be used for physical extrapolation. This is a consequence of the Bayesian approach that, one might argue, aims to build physical models.  Extrapolation models can potentially be used to generate quantum dynamics predictions in a wide range of Hamiltonian parameters, including where rigorous quantum calculations become prohibitively difficult.  Extrapolation models may also be used to accelerate the numerical integration of the Schr\"{o}dinger equation. This can be achieved by replacing some of the matrix inversions required for the numerical integration of coupled second-order differential equations with kernel-based models of the intermediate propagation results. However, much work remains to be done to understand the limitations of the extrapolation models. It is particularly important to explore if the extrapolation models can provide more accurate results than the approximate dynamical methods based on decoupling approximations and/or elimination or classical treatment of quantum degrees of freedom.


\section*{Acknowledgments}
The work of the author is supported by NSERC of Canada.

\end{document}